\documentclass[12pt]{article}
\usepackage{amssymb,amsmath,amsthm,amsfonts,amscd}
 \usepackage{graphicx}
\newcommand\be{\begin{equation}}
\newcommand\ee{\end{equation}}
\newcommand\bea{\begin{eqnarray}}
\newcommand\eea{\end{eqnarray}}
\newcommand\ket[1]{|#1\rangle}

\newcommand{\fatalpha}{{\bf \alpha \kern -0.44em \alpha}}
\newcommand{\fatsigma}{{\bf \sigma \kern -0.54em \sigma}}
\newcommand{\tpchi}{{\bf D \kern -0.35em D}}
\newcommand{\llambda}{{\bf \lambda \kern -0.45em \lambda}}

\renewcommand{\theequation}{\arabic{equation}}
\renewcommand{\theequation}{\thesection.\arabic{equation}}
\bibliography{plain}
\title{\bf Multi-qubit stabilizer and cluster entanglement witnesses
}\vspace{20mm}
\author{ M. A. Jafarizadeh $^{a,b,c}$
 \thanks{E-mail:jafarizadeh@tabrizu.ac.ir}  ,
 G. Najarbashi  $^{a,b}$
 \thanks{E-mail:najarbashi@tabrizu.ac.ir} ,
  Y. Akbari  $^{a,b}$
 \thanks{E-mail:y-akbari@tabrizu.ac.ir},
  H. Habibian $^{a}$
 \thanks{E-mail:hesam-habibian@sicatechdec.com}
\\ $^a${\small Department of Theoretical Physics and Astrophysics,
University of Tabriz, Tabriz 51664, Iran.} \\ $^b${\small
Institute for Studies in Theoretical Physics and Mathematics,
Tehran 19395-1795, Iran.} \\ $^c${\small Research Institute for
Fundamental Sciences, Tabriz 51664, Iran. }} \pagebreak

\vspace{20mm}
\begin{document}
\maketitle \vspace{15mm}
\newpage
\begin{abstract}
One of the problems concerning entanglement witnesses (EWs) is the
construction of them by a given set of operators. Here several
multi-qubit EWs called stabilizer EWs are constructed by using the
stabilizer operators of some given multi-qubit states such as GHZ,
cluster  and exceptional states. The general approach  to manipulate
the  multi-qubit stabilizer EWs by exact(approximate) linear
programming (LP) method  is described and it is shown that the
Clifford group  play a crucial role in finding the hyper-planes
encircling the feasible region. The optimality, decomposability and
non-decomposability of constructed stabilizer EWs are discussed.
\end{abstract}


{\bf PACs Index: 03.65.Ud}

\vspace{70mm}
\newpage
\section{Introduction}
The problem of characterizing n-qubit entanglement  has motivated
considerable interest in the literature. This problem was raised
within the context of quantum information and quantum computation
processes such as teleportation, dense coding and quantum key
distribution \cite{nielsen1,ekert1,preskill} which consider the
physical phenomenon  of entanglement as a resource. Though there
are a number of very useful and spectacular results  for detecting
the presence of entanglement in pure and mixed states of
multipartite systems, the subject is still at its infancy
\cite{horod1,lewen1,terhal1,bruss1}.
\par
Among the different criteria to analyze the separability  of
quantum states  the entanglement witnesses (EWs) are of special
interest since it has been proved that for any entangled state
there exists at least one EW detecting it \cite{woron1,horod2}.
The EWs are Hermitian operators which have non-negative
expectation values over all separable states and  detect some
entangled states. A great deal of investigation has been devoted
to the study of EWs, considering their decomposability, optimality
\cite{lewen2}, optimal setups for local measurements of witnesses
\cite{guhne2,guhne3} and even their use in the characterization of
entanglement in important physical systems \cite{toth1,bruk1,wu1}.
Inside the several problems concerning the EWs, the problem of how
to construct EWs by a given set of operators has a great
importance. From a different point of view, a very useful approach
to construct EWs is the linear programming (LP)
\cite{doherty1,doherty2,jafar1,jafar2,jafar3, multispinor}, a
special case of convex optimization which can be solved by using
very efficient algorithms such as the simplex and interior-point
methods ( see e.g. \cite{boyd,chong}). In fact, in order to a
hermitian operator $W$ be an EW, it must posses at least one
negative eigenvalue and the expectation value of $W$ over any
separable state must be non-negative. Therefore, for determination
of EWs, one needs to determine the minimum value of this
expectation value over the feasible region (the minimum value must
be non-negative) and hence the problem reduces to an optimization
over the convex set of feasible region. For example, in
\cite{jafar1,jafar2} the manipulation of generic Bell-states
diagonal EWs has been reduced to such an optimization problem. It
has been shown that, if the feasible region for this optimization
problem constructs a polygon by itself, the corresponding boundary
points of the convex hull will minimize exactly the optimization
problem. This problem is called linear programming (LP) and the
simplex method is the easiest way of solving it. If the feasible
region is not a polygon, with the help of tangent planes in this
region at points which are determined either analytically or
numerically, one can define a new convex hull which is a polygon
and has encircled the feasible region. The points on the boundary
of the polygon can approximately determine the minimum value of
the optimization problem. Thus the approximated value is obtained
via LP. In general, it is difficult to find this region and solve
the corresponding optimization problem; thus, it is difficult to
find any generic multipartite EW. Recently, in Ref. \cite{jafar3},
a new class of EWs called reduction type EWs has been introduced
for which the feasible regions turn out to be convex polygons.
Also, in Ref.\cite{multispinor}, some kinds of Bell-states
diagonal relativistic multispinor EWs have been constructed which
can be manipulated by using exact and approximate LP method.
\par
On the other hand, stabilizer formalism and Clifford group
operations have been proved to be useful in quantum error
correction (theory of stabilizer codes)
\cite{gott1,gott2,moor1,cleve1}, quantum computing, entanglement
distillation \cite{moor2,moor3,raus1} and entanglement detection
\cite{guhne2,guhne3,toth1}. In this paper, we link stabilizer
theory and Clifford group operations with structure of new type
EWs, the so-called stabilizer EWs (SEWs). As we will show all
vertex points and hyper-planes surrounding feasible regions (i.e.,
the regions
   coming from the positivity of EWs with separable states) can be obtained just from a few ones  by
applying the Clifford group operations. The optimality of SEWs
corresponding to hyper-planes surrounding feasible region is
discussed in detail and it is shown that the optimality has a
close connection with the common eigenvectors of stabilizer
operators.
\par
The paper is organized as follows: In Section 2, we review the
basic notions and definitions of EWs relevant to our study and
describe a general approach of constructing stabilizer EWs by
exact and approximate LP method. In Section 3, we consider the
construction of  SEWs that can be solved by exact LP method and as
instances of such SEWs, we describe the SEWs of GHZ and cluster
states in details and give a brief discussion about SEWs of five,
seven, eight and nine qubit stabilizer states. Also the role of
Clifford group operations is studied in this construction. Section
4 is devoted to an analysis of optimality of the introduced SEWs.
It is proved that some of the SEWs which correspond to surrounding
half-planes of the feasible regions are optimal. In Section 5, we
consider the decomposability or non-decomposability of GHZ and
cluster states SEWs and show that the three-qubit SEWs are all
decomposable but for more than three-qubit, there exist
non-decomposable SEWs as well. In Section 6, we give some
entangled mixed states that can be detected by the SEWs. Section 7
is devoted to construct SEWs that their feasible regions are not
polygons by themselves but can be approximated by  polygons and
then solved by LP method. The paper is ended with a brief
conclusion and two appendices.
\section{Stabilizer EWs and LP method}
\subsection{Entanglement witnesses}
First let us recall the definition of entanglement and
separability. An n-partite quantum mixed state $\rho\in
{\cal{B}}({\cal{H}})$ (the Hilbert space of bounded operators
acting on
$\mathcal{H}={\cal{H}}_{d_{1}}\otimes...\otimes{\cal{H}}_{d_{n}})$
is called fully separable if it can be written as a convex
combination of pure product states, that is
\begin{equation}\label{fullsep}
    \rho=\sum_{i} p_{i} | \alpha_{i}^{(1)} \rangle \langle \alpha_{i}^{(1)} |\otimes
    | \alpha_{i}^{(2)} \rangle \langle \alpha_{i}^{(2)}
    |\otimes...\otimes| \alpha_{i}^{(n)} \rangle \langle \alpha_{i}^{(n)} |
\end{equation}
where $|\alpha_{i}^{(j)}\rangle$ with $j=1,...,n$ are  arbitrary
but normalized vectors lying in the  $\mathcal{H}_{d_{j}}$, and
$p_{i}\geq0$ with $\sum_{i}p_{i}=1$. When this is not the case,
$\rho$ is called entangled. Although  the  definitions of
separable  and entangled states were extended to consider various
partitions of the original system into subsystems
\cite{dur1,brand1}, throughout the paper by separability  we mean
fully separability.
 \par
 An
entanglement witness $\mathcal{W}$ is a Hermitian operator such
that $Tr(\mathcal{W}\rho_{s})\geq 0 $ for all separable states
$\rho_{s}$ and there exists at least one entangled state
$\rho_{e}$ which can be detected by $\mathcal{W}$, that is $
Tr(\mathcal{W}\rho_{e})<0 $. Note that in the aforementioned
definition of EWs, we are not worry about the kind of entanglement
of the quantum state and we are rather looking for EWs which have
non-negative expectation values over all separable states despite
the fact that they have some negative eigenvalues. The existence
of an EW for any entangled state is a direct consequence of
Hahn-Banach theorem \cite{rudin1} and the fact that the set of
separable density operators is convex and closed.
\par
Based on the notion of partial transposition, the EWs are
classified into two classes: decomposable (d-EW) and
non-decomposable (nd-EW). An EW $\mathcal{W}$ is called
decomposable if there exist positive operators $\mathcal{P},
\mathcal{Q}_{K}$ such that
\begin{equation}
 \mathcal{W}=\mathcal{P}+\sum_{K\subset
 \mathcal{N}}\mathcal{Q}_{K}^{T_{K}}
\end{equation}
where $\mathcal{N}:=\{1,2,3,...,n\}$ and  $T_{K}$ denotes the
partial transposition with respect to partite $K\subset
\mathcal{N}$ and it is non-decomposable if it can not be written
in this form \cite{doherty1}. Clearly d-EW can not detect bound
entangled states (entangled states with positive partial transpose
(PPT) with respect to all subsystems) whereas there are some bound
entangled states which can be detected by an nd-EW.
\par
Usually one is interested in finding EWs $\mathcal{W}$ which
detect entangled states in an optimal way in the sense that when
we subtract any positive operator from $\mathcal{W}$, then it does
not remain an EW anymore \cite{lewen1}. In other words, if there
exist $\epsilon> 0$ and a positive operator $ \mathcal{P}$ such
that $\mathcal{W'}=\mathcal{W}-\epsilon \mathcal{P}$ is again an
EW, then we conclude that $\mathcal{W}$ is not optimal and
otherwise it is an optimal EW.
\subsection{Manipulation of EWs by exact and approximate LP method}
This subsection is devoted to  describe linear programming (LP)
and general approach  to manipulate the so-called stabilizer EWs
by exact or approximate LP method \cite{boyd}.
\par
Consider a non-positive Hermitian operator of the form
\begin{equation}\label{wgen}
\mathcal{W}=a_{_{0}}I+\sum_{i} a_{i}Q_{_{i}}
\end{equation}
where $Q_{_{i}}$ are Hermitian operators and $a_{_{i}}$'s are real
parameters with $a_{_{0}}>0$. In this work, the operators
$Q_{_{i}}$ will be considered as operations of a given multi-qubit
stabilizer group. The stabilizer operations are mutually commuting
and their eigenvalues are +1 and -1. We will attempt to choose the
real parameters $a_{i}$ such that $\mathcal{W}$ becomes an EW. To
this aim, we introduce the maps
\begin{equation}\label{varp}
    P_{_{i}}=Tr(Q_{_{i}}\rho_{s})
\end{equation}
for any separable state $\rho_{s}$. The maps $P_{_{i}}$ map the
convex set of separable states into a region which will be named
feasible region. Since $-1\leq P_{_{i}}\leq1$ for all $i$, the
feasible region is bounded and lies inside the hypercube defined
by $-1\leq P_{_{i}}\leq1$ for all $i$. The first property of an EW
is that its expectation value over any separable state is
non-negative, i.e., the condition
$$
\mathcal{F}_{_{\mathcal{W}}}:=Tr(\mathcal{W}\rho_{s})=a_{_{0}}+
 \sum_{i}a_{_{i}}P_{_{i}}\geq0
$$
is satisfied for any point of the feasible region. For satisfying
this condition, it is sufficient that the minimum value of
$\mathcal{F}_{_{\mathcal{W}}}$ be non-negative. Therefore, for
determination of EWs of type (\ref{wgen}), one needs to determine
the minimum value of $a_0+\sum_{i=1}^na_iP_i$ over the feasible
region (the minimum value must be non-negative) and hence the
problem reduces to the optimization of the linear function
$a_0+\sum_{i=1}^na_iP_i$ over the convex set of feasible region.

We note that, the quantity $\mathcal{F}_{_{\mathcal{W}}}$ achieves
its minimum value for pure product states, since every separable
mixed state $\rho_{s}$ can be written as a convex combination of
pure product states, say $\rho_{s}=\sum_{i}p_{i}
|\Upsilon_{i}\rangle\langle\Upsilon_{i}|$ with $p_{i}\geq0$ and
$\sum_{i}p_{i}=1$, whence we have
\begin{equation}
Tr(\mathcal{W}\rho_{s})=\sum_{i}p_{i}Tr(\mathcal{W}|\Upsilon_{i}\rangle\langle\Upsilon_{i}|)\geq
C_{min},
\end{equation}
$$
\mathrm{with}  \quad\quad C_{min}=\min_{_{|\Upsilon\rangle\in
D_{prod.}}} \;Tr(\mathcal{W} |\Upsilon\rangle\langle\Upsilon|)
$$
where, $D_{prod.}$ denotes the set of pure product states. In this
work, we are interested in the EWs that their feasible regions are
of simplex (or at most convex polygon) types. The manipulation of
these EWs  amounts to
\begin{equation}\label{lp}
\begin{array}{c}
\mathrm{minimize}\quad \mathcal{F}_{_{\mathcal{W}}}=a_{_{0}}+
\sum_{i}a_{_{i}}P_{_{i}}\\
\mathrm{subject\; to}\quad
\sum_{i}(c_{_{ij}}P_{_{i}}-d_{_{j}})\geq 0 \quad  j=1,2,...\\
\end{array}
\end{equation}
where $c_{_{ij}}$ and $ d_{_{j}}$  are parameters of hyper-planes
surrounding  the feasible regions. So the problem reduces to a LP
problem.  On the basis of LP method, minimum of an objective
function $\mathcal{F}_{_{\mathcal{W}}}$ always  occurs at the
vertices of bounded feasible region. Therefore the vertices of
feasible region come from pure product states.
\par
It is necessary to distinguish between two cases: (\textbf{a})
exactly soluble, and (\textbf{b}) approximately soluble EWs. In
the case \textbf{a}, the boundaries (constraints on $P_{_{i}}$)
come from finite vertices arising from pure product states and
construct a convex polygon, while in the case \textbf{b} the
feasible region is not a polygon and the boundaries may be bounded
convex hypersurfaces. In this case, with the help of tangent
planes in this region at points which are determined either
analytically or numerically, one can define a new convex hull
which is a polygon encircling the feasible region, i.e., we
approximate the boundaries with hyper-planes and clearly some
vertices do not arise from pure product states. The points on the
boundary of the polygon can approximately determine the minimum
value of $\mathcal{F}_{_{\mathcal{W}}}$ in (\ref{lp}). Thus the
approximated value is obtained via LP. The both cases can be
solved by the well-known simplex method. The simplex algorithm is
a common algorithm used to solve an optimization problem with a
polytope feasible region, such as a linear programming problem. It
is an improvement over the algorithm to test all feasible solution
of the convex feasible region and then choose the optimal feasible
solution. It does this by moving from one vertex to an adjacent
vertex, such that the objective function is improved. This
algorithm still guarantees that the optimal point will be
discovered. In addition, only in the worst case scenario will all
vertices be tested. Here, considering the scope of this paper, a
complete treatment of the simplex algorithm is unnecessary; for a
more complete treatment please refer to any LP text such as
\cite{boyd,chong}.
\section{Exactly soluble stabilizer EWs}\label{sec1}
In this section we consider the construction of  stabilizer EWs
(SEWs) which can be solved exactly by the LP method. In motivating
this construction, we begin with  EWs which can be constructed by
the stabilizer operations of the multi-qubit GHZ state.

But before proceeding, it should be noticed that the Hermitian
operator of the form (\ref{wgen}) can not be a SEW when all the
$Q_{_{i}}$'s  form  pairwise locally commuting set. Two operators
$$
Q=L_{1}\otimes...\otimes L_{n}\quad \mathrm{and} \quad
Q'=K_{1}\otimes... \otimes K_{n}.
$$
are called locally commuting if $[L_{i},K_{i}]=0\;$, for all
$i=1,2,...,n$. To prove this assertion, consider the following
operator
$$
\mathcal{W}=a\ I + b\ Q + c\ Q'.
$$
Because of the commutativity of $K_{i}$ and $L_{i}$ we have the
$$
L_{i}=\sum_{\nu_{i}}\lambda_{\nu_{i}}^{(i)}|\psi_{\nu_{i}}^{(i)}\rangle\langle
\psi_{\nu_{i}}^{(i)}|\qquad,\qquad
K_{i}=\sum_{\nu_{i}}\mu_{\nu_{i}}^{(i)}|\psi_{\nu_{i}}^{(i)}\rangle\langle
\psi_{\nu_{i}}^{(i)}|.
$$
which in turn imply that the operator $\mathcal{W}$ can be written
as
$$
\mathcal{W}=a\ I + b\ \bigotimes_{i=1}^n
\sum_{\nu_{i}}\lambda_{\nu_{i}}^{(i)}|\psi_{\nu_{i}}^{(i)}\rangle\langle
\psi_{\nu_{i}}^{(i)}| + c\
\bigotimes_{i=1}^n\sum_{\nu_{i}}\mu_{\nu_{i}}^{(i)}|\psi_{\nu_{i}}^{(i)}\rangle\langle
\psi_{\nu_{i}}^{(i)}|
$$
$$
=
\sum_{\nu_{1}}...\sum_{\nu_{n}}(a+b\lambda_{\nu_{1}}^{(1)}...\lambda_{\nu_{n}}^{(n)}+
c\mu_{\nu_{1}}^{(1)}...\mu_{\nu_{n}}^{(n)})\
\bigotimes_{i=1}^n|\psi_{\nu_{i}}^{(i)}\rangle\langle
\psi_{\nu_{i}}^{(i)}|
$$
Now if we want  $\mathcal{W}$ to be an EW then it must has
non-negative expectation values with all pure product states which
means that all eigenvalues
$(a+b\lambda_{\nu_{1}}^{(1)}...\lambda_{\nu_{n}}^{(n)}+
c\mu_{\nu_{1}}^{(1)}...\mu_{\nu_{n}}^{(n)})$ are non-negative,
hence $\mathcal{W}$ is a positive operator. Therefore, the SEWs
can be constructed from the set of stabilizer operators $Q_{_{i}}$
that at least one pair of them is  not locally commuting.
\par
Throughout the paper, the generators of stabilizer groups are
chosen according to the table of appendix I. Of course, this
choice is arbitrary and one can take other elements as generators.
By the method presented here we can construct SEWs (exactly or
approximately) for completely different stabilizer groups.
\subsection{GHZ stabilizer EWs}
We consider   even case of GHZ SEWs which lies in realm of exactly
soluble  LP problems. The odd case is discussed in appendix III. A
similar construction can be made based on other elements of the
GHZ stabilizer group.
 \subsubsection{Even case}
Let us consider a situation in which the Hermitian operator is
composed of all generators of GHZ stabilizer group together  with
all even terms
$S_{_{1}}^{(\mathrm{GHZ})}S_{_{2k}}^{(\mathrm{GHZ})}$ (the name
even  refer to the index $2k$) as follows
\begin{equation}\label{witghzev}
    \mathcal{W}_{_{GHZ}}^{(n)}=a_{_{0}}I_{_{2^n}}+\sum_{k=1}^n
a_{_{k}}S_{_{k}}^{(\mathrm{GHZ})}+\sum_{k=1}^{n'}a_{_{1,2k}}S_{_{1}}^{(\mathrm{GHZ})}S_{_{2k}}^{(\mathrm{GHZ})}
\quad , \quad n':=\left[\frac{n}{2}\right],
\end{equation}
where, $S_{_{k}}^{(\mathrm{GHZ})}$ for $k=1,...,n$ are given in
the table of the Appendix I and the reader is referred to that
appendix for an overview of the stabilizer formalism.  Due to the
commutativity of all GHZ stabilizer generators,
 it is easy to see that the eigenvalues of $\mathcal{W}_{_{GHZ}}^{(n)}$ are
\begin{equation}\label{eigghzev}
    a_{_{0}}+\sum_{k=1}^n (-1)^{i_{k}}
a_{_{k}}+\sum_{k=1}^{n'}(-1)^{i_{1}+ i_{2k}}a_{_{1,2k}}\quad
,\quad \forall \ (i_{1},i_{2},...,i_{n})\in \{0,1\}^{n}.
\end{equation}
Evidently, when all eigenvalues are positive the above operator is
positive; otherwise it may be a SEW.
\par
For a separable state $\rho_{s}$, the positivity of
$$
Tr(\mathcal{W}_{_{GHZ}}^{(n)}\rho_{s})\geq0
$$
implies the positivity of the objective function
\begin{equation}\label{objghzev}
    \mathcal{F}_{\mathcal{W}_{_{GHZ}}^{(n)}}=a_{_{0}}+\sum_{k=1}^n
a_{_{k}}P_{_{k}}+\sum_{k=1}^{n'}a_{_{1,2k}}P_{_{1,2k}}\geq0,
\end{equation}
where
$$
P_{_{k}}=Tr(S_{_{k}}^{(\mathrm{GHZ})}\rho_{s})\quad,\quad
P_{_{1,2k}}=Tr(S_{_{1}}^{(\mathrm{GHZ})}S_{_{2k}}^{(\mathrm{GHZ})}\rho_{s}),
$$
and all of the $P_{_{k}}$'s and $P_{_{1,2k}}$'s lie in the
interval $[-1,1]$. Furthermore, the operator
$\mathcal{W}_{_{GHZ}}^{(n)}$ must has at least one negative
eigenvalue to become a SEW. To reduce the problem to a LP one and
to determine the feasible region, we require to know the vertices,
namely the extreme points of the feasible region. Vertex points of
the feasible region come from pure product states. The
coordinates  of vertex points can take one of three values +1, -1
and 0. Regarding the above considerations,  the product vectors
and the vertex points of the feasible region coming from them are
 listed in
 table 1,
\begin{table}[h]
\renewcommand{\arraystretch}{1}
\addtolength{\arraycolsep}{-2pt}
$$
\begin{array}{|c|c|}\hline
\mathrm{Product \ state} &
(P_{2},P_{3},...,P_{n-1},P_{n},P_{1},P_{1,2},P_{1,4},...,P_{1,2n'-2},P_{1,2n'})
\\ \hline
  |\Psi^{\pm}\rangle & (0,0,...,0,0,\pm1,0,0,...,0,0) \\
  \Lambda_{_{1}} |\Psi^{\pm}\rangle & (0,0,...,0,0,0,\pm1,0,...,0,0) \\
  \vdots & \vdots \\
  \Lambda_{_{n'}} |\Psi^{\pm}\rangle & (0,0,...,0,0,0,0,0,...,0,\pm1) \\
  \hline
  \Xi_{i_{2},...,i_{n}}|\Psi^{+}\rangle & \left((-1)^{i_{_{2}}},
  (-1)^{i_{_{2}}+i_{_{3}}},...,(-1)^{i_{_{n-2}}+i_{_{n-1}}},(-1)^{i_{_{n-1}}+i_{_{n}}},0,0,0,...,0,0\right)
    \\ \hline
\end{array}
$$
\caption{The product vectors and coordinates of vertices  for
$\mathcal{W}_{_{GHZ}}^{(n)}$ . }\label{tab1}
\renewcommand{\arraystretch}{1}
\addtolength{\arraycolsep}{-3pt}
\end{table}
where
\begin{equation}\label{tab1'}
    \begin{array}{c}
 |\Psi^{\pm}\rangle=|x^\pm\rangle_{_{1}}|x^+\rangle_{_{2}}|x^+\rangle_{_{3}}...|x^+\rangle_{_{n}}\\
 \Lambda_{_{k}}=\left({M^{(2k-1)}}\right)^{\dagger} M^{(2k)}\quad k=1,2,...,n'\\
 \Xi_{i_{2},...,i_{n}}=(\sigma_{_{x}}^{(2)})^{i_{_{2}}}...(\sigma_{_{x}}^{(n)})^{i_{_{n}}}
 \bigotimes_{j=1}^n H^{(j)}\quad,\quad \forall\ (i_{2},i_{3},...,i_{n})\in \{0,1\}^{n-1}\\
  \end{array}
\end{equation}
and $|x^\pm\rangle$ are eigenvectors of $\sigma_{x}$ with
eigenvalues $\pm1$. Here $M^{(k)}$ and $H^{(k)}$ are the
phase-shift operator and Hadamard transform  acting on particle
$k$  respectively (see appendix I). One can easily check by direct
calculation that the convex hull of the points listed in table 1
is contained in the feasible region and form a $(n-1)
2^{n'+2}$-simplex with the following boundary hyper-planes
$$
|P_{_{1}}\pm P_{_{j}}+\sum_{k=1}^{n'}(-1)^{i_{k}}P_{_{1,2k}}|= 1
,\quad j=2,...,n ,\quad \forall \ (i_{1},i_{2},...,i_{n'})\in
\{0,1\}^{n'}.
$$
On the other hand, in appendix II it is shown that the feasible
region is also contained in this simplex, i.e., the feasible
region is exactly determined by the intersection of the
half-spaces
\begin{equation}\label{ineghzev}
|P_{_{1}}\pm P_{_{j}}+\sum_{k=1}^{n'}(-1)^{i_{k}}P_{_{1,2k}}|\leq
1.
\end{equation}
In fact the half-spaces (\ref{ineghzev}) come from the positivity
of the expectation values of the operators
$$
\begin{array}{c}
  I_{_{2^n}}+S_{_{1}}^{(\mathrm{GHZ})}\pm
S_{_{j}}^{(\mathrm{GHZ})}+\sum_{k=1}^{n'}(-1)^{i_{k}}S_{_{1,2k}}^{(\mathrm{GHZ})}  \\
  I_{_{2^n}}-S_{_{1}}^{(\mathrm{GHZ})}\pm
S_{_{j}}^{(\mathrm{GHZ})}-\sum_{k=1}^{n'}(-1)^{i_{k}}S_{_{1,2k}}^{(\mathrm{GHZ})}
  \\
\end{array}
\quad,\quad j=2,...,n \quad,\quad \forall \
(i_{1},i_{2},...,i_{n'})\in \{0,1\}^{n'}
$$
over pure product states. We note that it is not necessary to
consider all the above operators since one can obtain them just by
applying some elements of the Clifford group (see Appendix I) on
the $2n^{'}+n^{''}$ (compare with  $(n-1) 2^{n'+2}$) following
operators
\begin{equation}\label{}
    \begin{array}{c}
       I_{_{2^n}}\pm\big(S_{_{1}}^{(\mathrm{GHZ})}+ S_{_{2j}}^{(\mathrm{GHZ})}
       +\sum_{k=1}^{n'}S_{_{1,2k}}^{(\mathrm{GHZ})}\big)\quad\quad
j=1,...,n^{'} \\
       I_{_{2^n}}-S_{_{1}}^{(\mathrm{GHZ})}- S_{_{2j+1}}^{(\mathrm{GHZ})}-\sum_{k=1}^{n'}S_{_{1,2k}}^{(\mathrm{GHZ})}\quad\quad
j=1,...,n^{''}. \\
     \end{array}
\end{equation}
For example  we get the operator
$S=I_{_{2^n}}+S_{_{1}}^{(\mathrm{GHZ})}-
S_{_{2j}}^{(\mathrm{GHZ})}
       -S_{_{1,2j}}^{(\mathrm{GHZ})}+\sum_{k\neq j}^{n'}S_{_{1,2k}}^{(\mathrm{GHZ})}$ from
the  operator $S'=I_{_{2^n}}+S_{_{1}}^{(\mathrm{GHZ})}+
S_{_{2j}}^{(\mathrm{GHZ})}
+\sum_{k=1}^{n'}S_{_{1,2k}}^{(\mathrm{GHZ})}$ under conjugation
with the Clifford operation $\sigma_{{x}}^{(2j)}$, i.e.,
$$
S= \big(\sigma_{{x}}^{(2j)} \big) S'
\big(\sigma^{(2j)}_{{x}}\big)^{\dag}.
$$
Now the problem of finding a pre-SEW (a hermitian operator with
non-negative expectation value over any separable state) of the
form (\ref{witghzev}) is reduced to the LP problem
\begin{equation}\label{lpghzev}
\begin{array}{c}
\mathrm{minimize}\quad
\mathcal{F}_{\mathcal{W}_{_{GHZ}}^{(n)}}=a_{_{0}}+\sum_{k=1}^n
a_{_{k}}P_{_{k}}+\sum_{k=1}^{n'}a_{_{1,2k}}P_{_{1,2k}}\\
\mathrm{subject\; to}\quad |P_{_{1}}\pm
P_{_{j}}+\sum_{k=1}^{n'}(-1)^{i_{k}}P_{_{1,2k}}|\leq 1 ,\quad
j=2,...,n ,\quad \forall \ (i_{1},i_{2},...,i_{n'})\in
\{0,1\}^{n'}
\end{array}
\end{equation}
 On the basis of LP method, minimum of an objective function
always  occurs at the vertices of the bounded feasible region.
Hence, if we put the coordinates of the vertices (see table 1) in
the objective function (\ref{objghzev}) and require the
non-negativity of the objective function on all vertices, we get
the conditions
\begin{equation}\label{ineqparaghz}
 \begin{array}{c}
  a_{_{0}}>0 \quad,\quad a_{_{0}}\geq  |a_{_{1}}|\quad,\quad a_{_{0}}\geq\sum_{i=2}^n |a_{_{i}}|\\
  a_{_{0}}\geq |a_{_{1,2k}}| \quad\quad k=1,...,n'\\
 \end{array}
\end{equation}
for parameters $a_i$. Evidently, these conditions are sufficient
to ensure that the objective function is non-negative on the whole
of the feasible region. If we take $a_{_{0}}=(n-1)$,
$a_{_{k}}=-1$, for all $k=1,...,n$ and $a_{_{1,2k}}=0$, for all
$k=1,...,n'$, which fulfill all the conditions of Eq.
(\ref{ineqparaghz}), then we get the SEW stated in Eq. (21) of
Ref. \cite{guhne1}. Also by taking $a_{_{0}}=1$, $a_{_{1}}=-1$ and
$a_{_{m}}=a_{_{1,m}}=-1$ ($m\geq 2$ is even) we have the SEWs
stated in Eq. (21) of the mentioned reference as  special cases.
\par
Fixing $a_{_{0}}$ in the space of parameters, all of the $a_i$'s
lie inside the polygon defined by inequalities
(\ref{ineqparaghz}). Now in order that the operator of
Eq.(\ref{witghzev})  becomes non-negative, all of its eigenvalues
in (\ref{eigghzev}) must be non-negative. The intersection of
half-spaces arising from the non-negativity of the eigenvalues
form a polyhedron inside the aforementioned polygon. The
complement of this polyhedron in the polygon is the where that the
operator (\ref{witghzev}) is SEW and will be named the SEWs
region.
\par
We assert that the SEWs region is non-empty. To confirm this
assertion, we discuss the case that all parameters $a_i$ are
positive since the discussion for other cases can be easily come
from by replacing any parameter by its negative value (except
$a_{_{0}}$ which is always positive). Because of the symmetry
between the parameters $a_{_{1}}$ and $a_{_{2k}}\ $'s
$(k=1,...,n')$, we can assume without loss of generality that $
a_{_{2}}\geq a_{_{4}}\geq a_{_{6}}\geq...\geq a_{_{2n'}}$. With
this assumption, all of the $2^{n''}+n'$ eigenvalues (with
$n''=[\frac{n-1}{2}]$):
$$
\left\{\begin{array}{c}
  \hspace{-2cm}a_{_{0}}+a_{_{1}}+\sum_{j=1}^{n'} a_{_{2j}}+
\sum_{j=1}^{n''} (-1)^{i_{2j+1}} a_{_{2j+1}}+\sum_{k=1}^{n'}a_{_{1,2k}} \quad\quad \forall \
(i_{3},i_{5},...,i_{2n''+1})\in \{0,1\}^{n''}\\
  a_{_{0}}+a_{_{1}}-a_{_{2l}}+\sum_{l\neq k=1}^{n'}a_{_{1,2k}}+\sum_{j=2}^n
 a_{_{j}}\quad\quad l=1,...,n' \hspace{8cm} \\
\end{array}\right.
$$
are non-negative and each of the $2^{n}-(2^{n''}+n')$ remaining
ones can take negative values.
\par
For example, consider the Hermitian operator
\begin{equation}
    \mathcal{W}_{_{GHZ}}^{(2)}=a_{_{0}}I_{_{4}}+
a_{_{1}}S_{_{1}}^{(\mathrm{GHZ})}+a_{_{2}}S_{_{2}}^{(\mathrm{GHZ})}+
a_{_{1,2}}S_{_{1}}^{(\mathrm{GHZ})}S_{_{2}}^{(\mathrm{GHZ})}
\end{equation}
with the following eigenvalues
\begin{equation}\label{eigenghz2}
\begin{array}{c}
  \omega_{_{1}}=a_{_{0}}+a_{_{1}}+a_{_{2}}+a_{_{1,2}}\quad,\quad \omega_{_{2}}=a_{_{0}}+a_{_{1}}-a_{_{2}}-a_{_{1,2}}\\
  \omega_{_{3}}=a_{_{0}}-a_{_{1}}+a_{_{2}}-a_{_{1,2}}\quad,\quad\omega_{_{4}}=a_{_{0}}-a_{_{1}}-a_{_{2}}+a_{_{1,2}}\;. \\
\end{array}
\end{equation}
We need only to consider the product state
$|x^{+}\rangle|x^{+}\rangle$ corresponding to the vertex point
$(1,0,0)$ since the product states corresponding to the other
vertex points can be obtained by applying the Clifford operations
$H\otimes H$ , $M\otimes M$ and $\sigma_{z} \otimes I$ on this
product state. Putting the vertex points in
$Tr(\mathcal{W}_{_{GHZ}}^{(2)}|\Upsilon\rangle\langle\Upsilon|)\geq0$
yields
$$
a_{_{0}}\geq|a_{_{1}}|,\quad a_{_{0}}\geq|a_{_{2}}|,\quad
a_{_{0}}\geq|a_{_{12}}|\;.
$$
So in the parameters space, the allowed values of $a$'s lie inside
a cube with edge length $a_{_{0}}$. The intersection of
half-spaces $\omega_{_{i}}\geq0\;(i=1,..,4)$ is a polyhedron
inside the cube whose vertices coincide with four vertices of the
cube and contains just the positive operators; the remaining part
of the cube is the region of SEWs. On the other hand the variables
$P_{_{i}}$ lie in the interval $[-1,1]$ and form a cube in the
space of variables. The convex hull of vertex points lies inside
this cube and has the eight boundary half-spaces
\begin{equation}
\begin{array}{c}
  |P_{_{1}}\pm P_{_{2}}+P_{_{1,2}}|\leq1\quad,\quad |P_{_{1}}\pm P_{_{2}}-P_{_{1,2}}|\leq1\;.\\
\end{array}
\end{equation}
The above half-spaces  define the feasible region (see Fig.1).
Four of these half-spaces which correspond to the positive
operators
\begin{equation}\label{wpos1}
\begin{array}{c}
  ^{1}\mathcal{P}_{_{GHZ}}=
  \left(I_{_{4}}+S_{_{1}}^{(\mathrm{GHZ})}+S_{_{2}}^{(\mathrm{GHZ})}+S_{_{12}}^{(\mathrm{GHZ})}\right)
  =4\;|\psi_{_{00}}\rangle\langle\psi_{_{00}}|\\
  ^{2}\mathcal{P}_{_{GHZ}}=\sigma_{_{z}}^{(1)} (^{1}\mathcal{P}_{_{GHZ}}) \sigma_{_{z}}^{(1)}=
  I_{_{4}}-S_{_{1}}^{(\mathrm{GHZ})}+S_{_{2}}^{(\mathrm{GHZ})}-S_{_{12}}^{(\mathrm{GHZ})} \\
  ^{3}\mathcal{P}_{_{GHZ}}=\sigma_{_{x}}^{(1)} (^{1}\mathcal{P}_{_{GHZ}}) \sigma_{_{x}}^{(1)}=
  I_{_{4}}+S_{_{1}}^{(\mathrm{GHZ})}-S_{_{2}}^{(\mathrm{GHZ})}-S_{_{12}}^{(\mathrm{GHZ})} \\
  ^{4}\mathcal{P}_{_{GHZ}}=\sigma_{_{y}}^{(1)} (^{1}\mathcal{P}_{_{GHZ}}) \sigma_{_{y}}^{(1)}=
  I_{_{4}}-S_{_{1}}^{(\mathrm{GHZ})}-S_{_{2}}^{(\mathrm{GHZ})}+S_{_{12}}^{(\mathrm{GHZ})} \\
\end{array}
\end{equation}
are in one-one correspondence with four vertices of the cube in
parameter space which are the same as the vertices of polyhedron
formed by the positive operators.
\par
For the purpose of later use, we  introduce
\begin{equation}
\ket{\psi_{_{i_{1}i_{2}...i_{n}}}}=(\sigma_{z})^{i_{1}}\otimes
(\sigma_{x})^{i_{2}}\otimes ... \otimes
(\sigma_{x})^{i_{n}}\ket{\psi_{_{00...0}}},
\end{equation}
 where  $
\ket{\psi_{_{00...0}}}=\frac{1}{\sqrt{2}}(\ket{00...0}+\ket{11...1})
$ is the  n-qubit GHZ  state. As implied by the
Eq.\;({\ref{wpos1}}), the three last positive operators can be
obtained from the first one via the action of some operations of
the Clifford group. The other four boundary half-spaces which
correspond to the optimal d-EWs
\begin{equation}\label{wopt1}
\begin{array}{c}
  ^{1}\mathcal{W}_{_{GHZ}}^{(opt)}=I_{_{4}}-S_{_{1}}^{(\mathrm{GHZ})}-S_{_{2}}^{(\mathrm{GHZ})}-S_{_{12}}^{(\mathrm{GHZ})}
  =4(|\psi_{_{11}}\rangle\langle\psi_{_{11}}|)^{T_{1}}\\
  ^{2}\mathcal{W}_{_{GHZ}}^{(opt)}= \sigma_{_{x}}^{(1)}(^{1}\mathcal{W}_{_{GHZ}}^{(opt)}) \sigma_{_{x}}^{(1)}=
  I_{_{4}}-S_{_{1}}^{(\mathrm{GHZ})}+S_{_{2}}^{(\mathrm{GHZ})}+S_{_{12}}^{(\mathrm{GHZ})} \\
  ^{3}\mathcal{W}_{_{GHZ}}^{(opt)}=\sigma_{_{z}}^{(1)} (^{1}\mathcal{W}_{_{GHZ}}^{(opt)}) \sigma_{_{z}}^{(1)}=
  I_{_{4}}+S_{_{1}}^{(\mathrm{GHZ})}-S_{_{2}}^{(\mathrm{GHZ})}+S_{_{12}}^{(\mathrm{GHZ})} \\
  ^{4}\mathcal{W}_{_{GHZ}}^{(opt)}=\sigma_{_{y}}^{(1)} (^{1}\mathcal{W}_{_{GHZ}}^{(opt)}) \sigma_{_{y}}^{(1)}=
  I_{_{4}}+S_{_{1}}^{(\mathrm{GHZ})}+S_{_{2}}^{(\mathrm{GHZ})}-S_{_{12}}^{(\mathrm{GHZ})} \\
\end{array}
\end{equation}
are in one-one correspondence with the remaining four vertices of
the cube in parameters space. From Eq.\;({\ref{wopt1}}) we see
that the three last optimal d-EWs can be also obtained from the
first one via the action of some operations of the Clifford group.
So as we had in \cite{jafar1}, the operators corresponding to the
boundary planes are either optimal SEWs or positive operators. In
this case, all of the witnesses are d-EWs since we can write them
as a convex combination of an optimal d-EW and a positive operator
from its opposite positive boundary plane.
\subsection{Multi-qubit cluster EWs}
We continue with  EWs which can be constructed by the stabilizer
operators of the cluster state and again consider two  even and
odd cases of the cluster SEWs which lie in the realm of exact LP
problems (refer to appendix III for odd case).
\subsubsection{Even case}
Let us consider the following Hermitian operators
\begin{equation}\label{witclev}
    \mathcal{W}_{_{C}}^{(n)}=a_{_{0}}I_{_{2^n}}+\sum_{k=1}^{n'}
a_{_{2k}}S_{_{2k}}^{(\mathrm{C})}
+a_{_{2m-1}}S_{_{2m-1}}^{(\mathrm{C})}+a_{_{2m-1,2m}}S_{_{2m-1}}^{(\mathrm{C})}S_{_{2m}}^{(\mathrm{C})},\quad
m=2,...,\left[\frac{n+1}{2}\right]-1
\end{equation}
In addition to the above operators, one can consider other
Hermitian operators which differ from the above operators only in
the last terms, that is the last terms of them are
 $a_{_{2m-2,2m-1}}S_{_{2m-2}}^{(\mathrm{C})}S_{_{2m-1}}^{(\mathrm{C})}$
with $m=2,...,\left[\frac{n+1}{2}\right]$. However, we will
consider only the operators (\ref{witclev}) since the treatment is
the same for others.  Due to the commutativity of all cluster
stabilizer generators, it is easy to see that the eigenvalues of
$\mathcal{W}_{_{C}}^{(n)}$ are
\begin{equation}\label{eigenclev}
a_{_{0}}+\sum_{j=1}^{n'} (-1)^{i_{2j}}
a_{_{2j}}+(-1)^{i_{2m-1}}a_{_{2m-1}}+(-1)^{i_{2m-1}+i_{2m}}a_{_{2m-1,2m}}
\quad, \quad \forall \ (i_{1},i_{2},...,i_{n})\in \{0,1\}^{n}
\end{equation}
To reduce the problem to a LP one and determine the feasible
region, we require to know the vertices, namely the extreme points
of the feasible region. For a separable state $\rho_{s}$, the
non-negativity of
$$
Tr(\mathcal{W}_{_{C}}^{(n)}\rho_{s})\geq0
$$
implies the non-negativity of the objective function
\begin{equation}\label{objclev}
\mathcal{F}_{\mathcal{W}_{_{C}}^{(n)}}=a_{_{0}}+\sum_{k=1}^{n'}
a_{_{2k}}P_{_{2k}}
+a_{_{2m-1}}P_{_{2m-1}}+a_{_{2m-1,2m}}P_{_{2m-1,2m}}\quad,\quad
m=2,...,\left[\frac{n+1}{2}\right]-1
\end{equation}
where,
$$
P_{_{2k}}=Tr(S_{_{2k}}^{(\mathrm{C})}\rho_{s})\quad,\quad
P_{_{2m-1,2m}}=Tr(S_{_{2m-1}}^{(\mathrm{C})}S_{_{2m}}^{(\mathrm{C})}\rho_{s}),
$$
and all of the $P_{_{2k}}$'s and $P_{_{2m-1,2m}}$'s lie in the
interval $[-1,1]$. The product vectors and the vertex points of
the feasible region coming from these product vectors are listed
in table 2
\begin{table}[h]
\renewcommand{\arraystretch}{1.2}
\addtolength{\arraycolsep}{-2pt}
$$
\begin{array}{|c|c|}\hline
 \mathrm{Product\ state }& (P_{2},P_{4},...,P_{2m-4},P_{2m-2},P_{2m-1},P_{2m},P_{2m+2},..., P_{2n'},P_{2m-1,2m}) \\ \hline
  \Lambda_{_{i_{1},...,i_{n'}}}^{(ev)}|\Phi\rangle & \big((-1)^{i_{1}},(-1)^{i_{2}},...,
  (-1)^{i_{m-2}},(-1)^{i_{m-1}},0,(-1)^{i_{m}},(-1)^{i_{m+1}},...,(-1)^{i_{n'}},0\big) \\
   \hline
  {\Lambda'}_{_{i_{1},...,i_{n'}}}^{(ev)}|\Phi\rangle& \big((-1)^{i_{1}},(-1)^{i_{2}},...,
  (-1)^{i_{m-2}},0,\pm1,0,(-1)^{i_{m+1}},...,(-1)^{i_{n'}},0\big)\\
  \hline
  {\Lambda''}_{_{i_{1},...,i_{n'}}}^{(ev)}|\Phi\rangle & \big((-1)^{i_{1}},(-1)^{i_{2}},...,(-1)^{i_{m-2}},0,0,0,(-1)
  ^{i_{m+1}},...,(-1)^{i_{n'}},(-1)^{i_{m}}\big)\\
  \hline
\end{array}
$$
\caption{The product vectors and coordinates of vertices for
$\mathcal{W}_{_{C}}^{(n)}$.}
\renewcommand{\arraystretch}{1}
\addtolength{\arraycolsep}{3pt}
\end{table}

where
$$
\begin{array}{c}
 |\Phi\rangle=|z^+\rangle_{_{1}}|x^+\rangle_{_{2}}|z^+\rangle_{_{3}}|x^+\rangle_{_{4}}|z^+\rangle_{_{5}}...
 |x^+\rangle_{_{n-1}}|z^+\rangle_{_{n}} \\
  \Lambda_{_{i_{1},...,i_{n'}}}^{(ev)}=\bigotimes_{j=1}^{n'}\left(\sigma_{_{z}}^{(2j)}\right)^{i_{j}}
 \quad,\quad \forall\ (i_{1},i_{2},...,i_{n'})\in \{0,1\}^{n'} \\
   {\Lambda'}_{_{i_{1},...,i_{n'}}}^{(ev)}= \Lambda_{_{i_{1},...,i_{n'}}}^{(ev)}H^{(2m-2)}H^{(2m-1)}H^{(2m)} \\
   {\Lambda''}_{_{i_{1},...,i_{n'}}}^{(ev)}=
   \Lambda_{_{i_{1},...,i_{n'}}}^{(ev)}H^{(2m-2)}M^{(2m-1)}H^{(2m-1)}M^{(2m)}\\
\end{array}
$$
For a given $m$, the convex hull of the above vertices, the
feasible region, is a $(2n'+12)$-simplex formed by the
intersection of the following half-spaces
\begin{equation}\label{ineclev}
\begin{array}{c}
  |P_{2m-1}\pm P_{2m-2}+P_{2m-1,2m}|\leq1 \\
  |P_{2m-1}\pm P_{2m-2}-P_{2m-1,2m}|\leq1 \\
  |P_{2m-1}\pm P_{2m}+P_{2m-1,2m}|\leq1 \\
  |P_{2m-1}\pm P_{2m}-P_{2m-1,2m}|\leq1 \\
  \hspace{1cm}|P_{2k}|\leq1 \quad,\quad m,m-1\neq k=1,...,n'\\
\end{array}
\end{equation}
(see Appendix II). In fact the half-spaces (\ref{ineclev}) come
from the non-negativity of the expectation values of their
corresponding operators
$$
\begin{array}{c}
I+S_{_{2m-1}}^{(\mathrm{C})}\pm S_{_{2m-2}}^{(\mathrm{C})}
+S_{_{2m-1}}^{(\mathrm{C})}S_{_{2m}}^{(\mathrm{C})}\quad,\quad
I-S_{_{2m-1}}^{(\mathrm{C})}
\mp S_{_{2m-2}}^{(\mathrm{C})}-S_{_{2m-1}}^{(\mathrm{C})}S_{_{2m}}^{(\mathrm{C})} \\
I+S_{_{2m-1}}^{(\mathrm{C})}\pm S_{_{2m-2}}^{(\mathrm{C})}
-S_{_{2m-1}}^{(\mathrm{C})}S_{_{2m}}^{(\mathrm{C})}\quad,\quad
I-S_{_{2m-1}}^{(\mathrm{C})}\mp S_{_{2m-2}}^{(\mathrm{C})}+S_{_{2m-1}}^{(\mathrm{C})}S_{_{2m}}^{(\mathrm{C})}\\
I+S_{_{2m-1}}^{(\mathrm{C})}\pm S_{_{2m}}^{(\mathrm{C})}
+S_{_{2m-1}}^{(\mathrm{C})}S_{_{2m}}^{(\mathrm{C})}\quad,\quad
I-S_{_{2m-1}}^{(\mathrm{C})}
\mp S_{_{2m}}^{(\mathrm{C})}-S_{_{2m-1}}^{(\mathrm{C})}S_{_{2m}}^{(\mathrm{C})}\\
I+S_{_{2m-1}}^{(\mathrm{C})}\pm S_{_{2m}}^{(\mathrm{C})}
-S_{_{2m-1}}^{(\mathrm{C})}S_{_{2m}}^{(\mathrm{C})}\quad,\quad
I-S_{_{2m-1}}^{(\mathrm{C})}\mp S_{_{2m}}^{(\mathrm{C})}+S_{_{2m-1}}^{(\mathrm{C})}S_{_{2m}}^{(\mathrm{C})}\\
I\pm S_{_{2k}}^{(\mathrm{C})}  \quad,\quad m,m-1\neq k=1,...,n'\\
\end{array}
$$
over pure product states. We note that it is not necessary to
consider all the above operators, since one can obtain them just
by applying some elements of the Clifford group on the $4$
(compare with  $2n'+12$) following operators
\begin{equation}
\begin{array}{c}
I\pm S_{_{2m-1}}^{(\mathrm{C})}\pm S_{_{2m}}^{(\mathrm{C})}\pm
S_{_{2m-1}}^{(\mathrm{C})}S_{_{2m}}^{(\mathrm{C})}\\
I-S_{_{2m-1}}^{(\mathrm{C})}-S_{_{2m-2}}^{(\mathrm{C})}-S_{_{2m-1}}^{(\mathrm{C})}S_{_{2m}}^{(\mathrm{C})}\\
I-S_{_{2}}^{(\mathrm{C})}\\
\end{array}
\end{equation}
For instance, the Clifford operation
$$
U=(CN_{42})(CN_{53})(CN_{13})(CN_{24})\in Cl(n)
$$
transforms $S_{_{2}}^{(\mathrm{C})}$ to $S_{_{4}}^{(\mathrm{C})}$
by conjugation, i.e.,
$$
U S_{_{2}}^{(\mathrm{C})} U^\dagger=S_{_{4}}^{(\mathrm{C})}
$$
Now the problem of finding a pre-SEW of the form (\ref{witclev})
is reduced to a LP problem with objective function (\ref{objclev})
and constraints (\ref{ineclev}). If we put the coordinates of
vertices (see table 2) in the objective function (\ref{objclev})
and require the non-negativity of the objective function on all
vertices we get the conditions
\begin{equation}\label{ineqparaclev}
    \begin{array}{c}
  a_{_{0}} \geq \sum_{j=1}^{n'}|a_{_{2j}}| \\
  a_{_{0}}\geq\sum_{j=1}^{m-2}|a_{_{2j}}|+\sum_{j=m+1}^{n'}|a_{_{2j}}|+|a_{_{2m-1}}|\\
  a_{_{0}}\geq\sum_{j=1}^{m-2}|a_{_{2j}}|+\sum_{j=m+1}^{n'}|a_{_{2j}}|+|a_{_{2m-1,2m}}|\\
\end{array}
\end{equation}
for the parameters $a_i$. Evidently, these conditions are
sufficient to ensure that the objective function is non-negative
on the whole of the feasible region. Cluster SEWs (\ref{witclev})
and the odd case discussed in appendix III contain the SEWs in
Eqs. (36) and (37) of Ref. \cite{guhne1} as special cases.
\par
Fixing $a_{_{0}}$ in the space of parameters, all of the $a$'s lie
inside the polygon defined by inequalities (\ref{ineqparaclev}).
Now in order that the operator of Eq.(\ref{witclev})  becomes
positive, all of its eigenvalues in (\ref{eigenclev}) must be
non-negative. The intersection of half-spaces arising from the
non-negativity of eigenvalues form a polyhedron inside the
aforementioned polygon. The same reasoning as in the even case of
GHZ SEWs, shows that the SEWs region is non-empty. For example,
consider the operator
$$
\mathcal{W}_{_{Cl}}^{(4)}=a_{_{0}}I_{_{2^4}}+
a_{_{1}}S_{_{1}}^{(\mathrm{C})}+a_{_{2}}S_{_{2}}^{(\mathrm{C})}+a_{_{4}}S_{_{4}}^{(\mathrm{C})}+
a_{_{1,2}}S_{_{1}}^{(\mathrm{C})}S_{_{2}}^{(\mathrm{C})}.
$$
The eigenvalues of this operator are
$$
\begin{array}{c}
  \omega_{_{1}}=a_{_{0}}+a_{_{1}}+a_{_{2}}+a_{_{4}}+a_{_{1,2}}\quad,\quad
  \omega_{_{2}}=a_{_{0}}+a_{_{1}}-a_{_{2}}+a_{_{4}}-a_{_{1,2}} \\
  \omega_{_{3}}=a_{_{0}}+a_{_{1}}+a_{_{2}}-a_{_{4}}+a_{_{1,2}}\quad,\quad
  \omega_{_{4}}=a_{_{0}}+a_{_{1}}-a_{_{2}}-a_{_{4}}-a_{_{1,2}} \\
  \omega_{_{5}}=a_{_{0}}-a_{_{1}}+a_{_{2}}+a_{_{4}}-a_{_{1,2}}\quad,\quad
  \omega_{_{6}}=a_{_{0}}-a_{_{1}}+a_{_{2}}-a_{_{4}}-a_{_{1,2}}\\
  \omega_{_{7}}=a_{_{0}}-a_{_{1}}-a_{_{2}}+a_{_{4}}+a_{_{1,2}}\quad,\quad
  \omega_{_{8}}=a_{_{0}}-a_{_{1}}-a_{_{2}}-a_{_{4}}+a_{_{1,2}}\\
\end{array}
$$
Without loss of generality we can assume that $a_{_{1}}\geq
a_{_{2}}$. With this assumption, the  first four eigenvalues
$\omega_{_{1}},\omega_{_{2}},\omega_{_{3}}$ and $\omega_{_{4}}$
are always positive. Now let $\omega_{_{5}}$ and $\omega_{_{6}}$
be negative, i.e., $a_{_{0}}+a_{_{2}}<a_{_{1}}+a_{_{1,2}}$. In
this case, $\omega_{_{7}}$ and $\omega_{_{8}}$ can not be negative
and vice versa. Therefore with these considerations, among the
eight eigenvalues only the pair $\omega_{_{5}},\omega_{_{6}}$ or
$\omega_{_{7}},\omega_{_{8}}$ can be negative. The explicit form
of some four-qubit cluster SEWs is postponed to section
\ref{decsec}.
\section{Optimality  of SEWs}\label{optsec}
Another advantage of stabilizer  EWs is that the optimality of the
EWs corresponding to the boundary hypereplanes of feasible region
can be easily determined by a simple method presented here.
Consider an EW corresponding to one of the hyper-planes in which
three terms $S_{_{i}}$, $S_{_{j}}$ and $S_{_{i}}S_{_{j}}$ appear
simultaneously such as
\begin{equation}\label{woptim}
    \mathcal{W}=I+(-1)^{i_{1}}S_{_{i}}+(-1)^{i_{2}}S_{_{j}}+(-1)^{i_{3}}S_{_{i}}S_{_{j}}+...
    \quad\forall\ i_{1},i_{2},i_{3},...\in \{0,1\}.
\end{equation}
If there exist $\epsilon> 0$ and a positive operator $
\mathcal{P}=|\psi\rangle\langle\psi|,$ such that
$\mathcal{W'}=\mathcal{W}-\epsilon |\psi\rangle\langle\psi|$ is
again an EW then we conclude that $\mathcal{W}$ is not optimal,
otherwise it is. Note that there is no restriction in taking
$\mathcal{P}$ as a pure positive operator since every positive
operator can be expressed as a sum of pure positive operators with
positive coefficients, i.e., $\mathcal{P}=\sum_{i}\lambda_{_{i}}
|\psi_{_{i}}\rangle\langle\psi_{_{i}}|$ with all
$\lambda_{_{i}}\geq0$. If $\mathcal{W'}$  be an  EW, then
$|\psi\rangle$ has to satisfy the constraint
$Tr(|\psi\rangle\langle\psi|\Upsilon\rangle\langle\Upsilon|)=|\langle\psi|\Upsilon\rangle|^2=0$
for any pure product state $|\Upsilon\rangle$ satisfying
$Tr(\mathcal{W}|\Upsilon\rangle\langle\Upsilon|)=0$. In other
words, $|\psi\rangle$ has to be orthogonal to all such pure
product states.
\par
Since in SEWs of the form (\ref{woptim}) considered so far there
is no pair of locally commuting operators, it is always possible
to find pure product vectors $|\Upsilon\rangle$ for which one of
the relations
\begin{equation}
\begin{array}{c}
  S_{_{i}}|\Upsilon\rangle=
(-1)^{i_{1}+1}|\Upsilon\rangle \\
  S_{_{j}}|\Upsilon\rangle=
(-1)^{i_{2}+1}|\Upsilon\rangle  \\
  S_{_{i}}S_{_{j}}|\Upsilon\rangle=
(-1)^{i_{3}+1}|\Upsilon\rangle  \\
\end{array}
\end{equation}
hold. The expectation value of $\mathcal{W}$ over such
$|\Upsilon\rangle$'s is zero and $|\psi\rangle$ cannot contain
such pure product vectors. All the eigenvectors of a stabilizer
operation can be chosen as pure product vectors, half of them with
eigenvalue +1 and the other half with eigenvalue -1, such that the
expectation value of other stabilizer operations over them be
zero. Because of Hermiticity of stabilizer operations, their
eigenvectors can be used as a basis.
\par
Let us assume that $|\Upsilon_{k}\rangle$'s are pure product
eigenvectors of $S_{_{i}}$ with eigenvalues $(-1)^{i_{1}+1}$ and
$|\Upsilon_{k}^{\perp}\rangle$'s are its pure product eigenvectors
with eigenvalues $(-1)^{i_{1}}$ that have been chosen according to
the above prescription. So the expectation value of $\mathcal{W}$
over $|\Upsilon_{k}\rangle$'s is zero and $|\psi\rangle$ cannot
contain $|\Upsilon_{k}\rangle$'s that is
$|\psi\rangle=\sum_{k}|\Upsilon_{k}^{\perp}\rangle$. This implies
that $ S_{_{i}}|\psi\rangle= (-1)^{i_{1}}|\psi\rangle$. By the
same reasoning we conclude that $ S_{_{j}}|\psi\rangle=
(-1)^{i_{2}}|\psi\rangle$ and $ S_{_{i}}S_{_{j}}|\psi\rangle=
(-1)^{i_{3}}|\psi\rangle$. On the other hand, we have $
S_{_{i}}S_{_{j}}|\psi\rangle=(-1)^{i_{2}}S_{_{i}}|\psi\rangle=
(-1)^{i_{1}+i_{2}}|\psi\rangle$. Hence, if $i_{3}\neq
i_{1}+i_{2}$, i.e., if $i_{3}=i_{1}+i_{2}+1$, we get into a
contradiction and $\mathcal{W}$ is optimal. Therefore,  among all
SEWs of the form (\ref{woptim}) the following ones are optimal
\begin{equation}\label{optgenform}
 \mathcal{W}_{opt}=I+(-1)^{i_{1}}S_{_{i}}+(-1)^{i_{2}}S_{_{j}}+(-1)^{i_{1}+i_{2}+1}S_{_{i}}S_{_{j}}+...
 \quad \quad\forall\ (i_{1},i_{2},...)\in \{0,1\}^{m}.
\end{equation}
With the same reasoning as above one can conclude that any SEW of
the general form
\begin{equation}\label{}
\mathcal{W}=I+(-1)^{i_{1}}S_{_{i}}+(-1)^{i_{2}}S_{_{j}}+(-1)^{i_{3}}S_{_{i}}S_{_{k}}+...
\quad \quad\forall\ i_{1},i_{2},...\in \{0,1\}.
\end{equation}
with $j\neq k$ is not optimal.
\par
For instance, in the case of the three-qubit $GHZ$-state,
\begin{equation}
\mathcal{W}_{_{GHZ}}^{(3)}=a_{_{0}}I_{_{8}}+
a_{_{1}}S_{_{1}}^{(\mathrm{GHZ})}+a_{_{2}}S_{_{2}}^{(\mathrm{GHZ})}+a_{_{3}}S_{_{3}}^{(\mathrm{GHZ})}+
a_{_{1,2}}S_{_{1}}^{(\mathrm{GHZ})}S_{_{2}}^{(\mathrm{GHZ})}
\end{equation}
the boundary half-spaces of the feasible region are
\begin{equation}\label{ineq3}
  \begin{array}{c}
    |P_{_{1}}\pm P_{_{j}}+P_{_{1,2}}|\leq 1\quad,\quad |P_{_{1}}\pm P_{_{j}}-P_{_{1,2}}|\leq 1\quad\quad j=2,3\\
  \end{array}
\end{equation}
Using Clifford group operations, we can obtain all of these
half-spaces only from the three half-spaces
\begin{equation}\label{}
  \begin{array}{c}
   |P_{_{1}}+ P_{_{2}}+P_{_{1,2}}|\leq 1\quad,\quad P_{_{1}}+ P_{_{3}}+P_{_{1,2}}\leq 1\\
  \end{array}
\end{equation}
The operators corresponding to the above boundary half-spaces are
\begin{equation}\label{bopghz3}
\begin{array}{c}
\mathcal{Q}_{_{GHZ}}=I_{_{8}}+
S_{_{1}}^{(\mathrm{GHZ})}+S_{_{2}}^{(\mathrm{GHZ})}+
S_{_{1}}^{(\mathrm{GHZ})}S_{_{2}}^{(\mathrm{GHZ})}=4\big(|\psi_{_{000}}\rangle\langle\psi_{_{000}}|
+|\psi_{_{001}}\rangle\langle\psi_{_{001}}|\big) \\
^{1}\mathcal{W}_{_{GHZ}}=I_{_{8}}-S_{_{1}}^{(\mathrm{GHZ})}-S_{_{2}}^{(\mathrm{GHZ})}-
S_{_{1}}^{(\mathrm{GHZ})}S_{_{2}}^{(\mathrm{GHZ})}=4\big(|\psi_{_{110}}\rangle\langle\psi_{_{110}}|
+|\psi_{_{111}}\rangle\langle\psi_{_{111}}|\big)^{T_{2}} \\
^{2}\mathcal{W}_{_{GHZ}}=I_{_{8}}-S_{_{1}}^{(\mathrm{GHZ})}-S_{_{3}}^{(\mathrm{GHZ})}-
S_{_{1}}^{(\mathrm{GHZ})}S_{_{2}}^{(\mathrm{GHZ})}=4|\psi_{_{101}}\rangle\langle\psi_{_{101}}|
+4\big(|\psi_{_{110}}\rangle\langle\psi_{_{110}}|\big)^{T_{2}}. \\
\end{array}
\end{equation}
It is seen that, in agreement with the above argument,
$^{1}\mathcal{W}_{_{GHZ}}$ is an optimal SEW but
$^{2}\mathcal{W}_{_{GHZ}}$ is not. Also, for the three-qubit
cluster state,
\begin{equation}
\mathcal{W'}_{_{C}}^{(3)}=a_{_{0}}I_{_{8}}+
a_{_{1}}S_{_{1}}^{(\mathrm{C})}+a_{_{2}}S_{_{2}}^{(\mathrm{C})}+a_{_{3}}S_{_{3}}^{(\mathrm{C})}+
a_{_{1,2}}S_{_{1}}^{(\mathrm{C})}S_{_{2}}^{(\mathrm{C})}
\end{equation}
using Clifford group operations, we can obtain all of the boundary
half-spaces only from the three half-spaces
$$
|P_{_{1}}+ P_{_{2}}+P_{_{1,2}}|\leq 1\quad,\quad P_{_{2}}+
P_{_{3}}+P_{_{1,2}}\leq 1
$$
and the operators corresponding to the above boundary half-spaces
are
\begin{equation}\label{localghzcluster}
    \begin{array}{c}
       H^{(1)}H^{(3)}\mathcal{Q}_{_{GHZ}}H^{(1)}H^{(3)}=I_{_{8}}+
       S_{_{1}}^{(\mathrm{C})}+S_{_{2}}^{(\mathrm{C})}+
       S_{_{1}}^{(\mathrm{C})}S_{_{2}}^{(\mathrm{C})} \\
       H^{(1)}H^{(3)}
       {^{1}\mathcal{W}_{_{GHZ}}}H^{(1)}H^{(3)}=I_{_{8}}-S_{_{1}}^{(\mathrm{C})}-S_{_{2}}^{(\mathrm{C})}-
       S_{_{1}}^{(\mathrm{C})}S_{_{2}}^{(\mathrm{C})} \\
       H^{(1)}H^{(3)}{^{2}\mathcal{W}_{_{GHZ}}}H^{(1)}H^{(3)}=I_{_{8}}-S_{_{2}}^{(\mathrm{C})}-S_{_{3}}^{(\mathrm{C})}-
       S_{_{1}}^{(\mathrm{C})}S_{_{2}}^{(\mathrm{C})} \\
     \end{array}
\end{equation}
Clearly, local unitary operations $U_{local}$ do not change the
optimality of EWs under the conjugation action  such as
$U_{local}W_{op}U_{local}^\dag$, hence among the above operators,
the second one remains optimal while the third one remains
non-optimal.
\section{Decomposability of SEWs}\label{decsec}
Another interesting feature of EWs which is necessary to study
about SEWs is decomposability. As it is well-known  that every
two-qubit EW is decomposable \cite{woron1,horod2,peres1}, we
discuss the three-qubit systems or more.
\subsection{Decomposability of $\mathcal{W}_{_{GHZ}}^{(n)}$}
First  consider three-qubit GHZ SEWs. The  inequalities
(\ref{ineqparaghz}) show that in the space of parameters all GHZ
SEWs lie inside the hypercube (again by fixing $a_{_{0}}$) but
this statement does not mean that any point of the region inside
the hypercube is an SEW. The region defined by the inequalities
\begin{equation}\label{ineq1}
    a_{_{0}}+ (-1)^{i_{1}}a_{_{1}}+(-1)^{i_{2}} a_{_{2}}+
    (-1)^{i_{3}} a_{_{3}}+(-1)^{i_{1}+i_{2}} a_{_{1,2}}\geq 0 \quad (i_{1},i_{2},i_{3})\in\{0,1\}^3
\end{equation}
is the place inside the hypercube where the operator
$\mathcal{W}_{_{GHZ}}^{(3)}$ have just positive eigenvalues and
hence is positive. First we consider the decomposability or
non-decomposability of SEWs lying over the edges of the hypercube.
These SEWs come from $^{1}\mathcal{W}_{_{GHZ}}$ and
$^{2}\mathcal{W}_{_{GHZ}}$ of (\ref{bopghz3}) by Clifford
operations. The $^{1}\mathcal{W}_{_{GHZ}}$ and SEWs coming from it
are optimal decomposable since their partial transpositions with
respect to some particles are positive.
\par
Now in the space of parameters $a$, we consider the coordinates of
points as $(a_{_{1}},a_{_{2}},a_{_{3}},a_{_{1,2}})$. Putting the
following four points (which lie over the edges of hypercube) in
$\mathcal{W}_{_{GHZ}}^{(3)}$ gives the following optimal SEWs
\begin{equation}\label{ineq2}
    \begin{array}{c}
  (1,1,0,-1)\  \rightarrow\ I_{_{8}}+S_{_{1}}^{(\mathrm{GHZ})}+S_{_{2}}^{(\mathrm{GHZ})}-
  S_{_{1}}^{(\mathrm{GHZ})}S_{_{2}}^{(\mathrm{GHZ})} \\
  (-1,1,0,1) \ \rightarrow\ I_{_{8}}-S_{_{1}}^{(\mathrm{GHZ})}+S_{_{2}}^{(\mathrm{GHZ})}+
 S_{_{1}}^{(\mathrm{GHZ})}S_{_{2}}^{(\mathrm{GHZ})} \\
  (1,-1,0,1) \ \rightarrow \ I_{_{8}}+S_{_{1}}^{(\mathrm{GHZ})}-S_{_{2}}^{(\mathrm{GHZ})}+
 S_{_{1}}^{(\mathrm{GHZ})}S_{_{2}}^{(\mathrm{GHZ})} \\
  (-1,-1,0,1)\ \rightarrow \ I_{_{8}}-S_{_{1}}^{(\mathrm{GHZ})}-S_{_{2}}^{(\mathrm{GHZ})}-
 S_{_{1}}^{(\mathrm{GHZ})}S_{_{2}}^{(\mathrm{GHZ})} \\
\end{array}
\end{equation}
The above SEWs are optimal decomposable since their partial
transpositions with respect to some particles are positive . A
convex cone which may be formed by connecting every four points of
Eq. (\ref{ineq2}) to its opposite positive hyper-plane in Eq.
(\ref{ineq1}) is d-SEWs. Note that the remaining operators in Eq.
(\ref{ineq3}) coming from some points in the space of parameters
are either  d-SEW or positive. Therefore we conclude that all the
three-qubit GHZ stabilizer EWs are decomposable. The discussion
for more than three-qubit is rather complicated. It is clear that
every EW with positive partial transpose with respect to some
particles is decomposable. Therefore  imposing the condition
\begin{equation}\label{}
    a_{_{0}}+ \sum_{k=1}^n (-1)^{i_{k}}a_{_{k}}+\sum_{k\in \mathcal{B}}(-1)^{i_{1}+i_{k}+1} a_{_{1,k}}+
    \sum_{k\in \mathcal{A}\setminus \mathcal{B}}(-1)^{i_{1}+i_{k}} a_{_{1,k}}\geq 0 \quad (i_{1},...,i_{n})\in\{0,1\}^n
\end{equation}
which in turn implies $\mathcal{W}^{T_{\mathcal{B}}}\geq 0$,
yields  the GHZ  decomposable SEWs where the $\mathcal{B}$ is any
nonempty subset of the set $\mathcal{A}=\{2,4,...,2n'\}$. Here
taking partial transpose with respect to the particles $2j$ and
$2j-1$ with $j=1,...,n'$ leads to the same result.
\par
In order to show that the $\mathcal{W}_{_{GHZ}}^{(n)}$ for $n\geq
4$ contain some nd-EWs, we discuss the four-qubit case in detail.
From (\ref{witghzev}), we have
$$
\mathcal{W}_{_{GHZ}}^{(4)}=a_{_{0}}I_{_{2^4}}+\sum_{k=1}^4
a_{_{k}}S_{_{k}}^{(\mathrm{GHZ})}+\sum_{k=1}^{2}a_{_{1,2k}}S_{_{1}}^{(\mathrm{GHZ})}S_{_{2k}}^{(\mathrm{GHZ})}
$$
Using the local Clifford operations, all the 48 Hermitian
operators corresponding to boundary half-spaces of the feasible
region can be obtained only from the following 5 ones
\begin{equation}\label{fouqubitGHZ}
    \begin{array}{c}
       ^{1}\mathcal{W}_{_{GHZ}}^{(4)}=I_{_{16}}+S_{_{1}}^{(\mathrm{GHZ})}+S_{_{2}}^{(\mathrm{GHZ})}+
  S_{_{1}}^{(\mathrm{GHZ})}S_{_{2}}^{(\mathrm{GHZ})}+S_{_{1}}^{(\mathrm{GHZ})}S_{_{4}}^{(\mathrm{GHZ})}\\
       ^{2}\mathcal{W}_{_{GHZ}}^{(4)}=I_{_{16}}+S_{_{1}}^{(\mathrm{GHZ})}+S_{_{4}}^{(\mathrm{GHZ})}+
  S_{_{1}}^{(\mathrm{GHZ})}S_{_{2}}^{(\mathrm{GHZ})}+S_{_{1}}^{(\mathrm{GHZ})}S_{_{4}}^{(\mathrm{GHZ})}\\
       ^{3}\mathcal{W}_{_{GHZ}}^{(4)}=I_{_{16}}-S_{_{1}}^{(\mathrm{GHZ})}-S_{_{2}}^{(\mathrm{GHZ})}-
  S_{_{1}}^{(\mathrm{GHZ})}S_{_{2}}^{(\mathrm{GHZ})}-S_{_{1}}^{(\mathrm{GHZ})}S_{_{4}}^{(\mathrm{GHZ})}\\
       ^{4}\mathcal{W}_{_{GHZ}}^{(4)}=I_{_{16}}-S_{_{1}}^{(\mathrm{GHZ})}-S_{_{4}}^{(\mathrm{GHZ})}-
  S_{_{1}}^{(\mathrm{GHZ})}S_{_{2}}^{(\mathrm{GHZ})}-S_{_{1}}^{(\mathrm{GHZ})}S_{_{4}}^{(\mathrm{GHZ})}\\
       ^{5}\mathcal{W}_{_{GHZ}}^{(4)}=I_{_{16}}-S_{_{1}}^{(\mathrm{GHZ})}-S_{_{3}}^{(\mathrm{GHZ})}-
  S_{_{1}}^{(\mathrm{GHZ})}S_{_{2}}^{(\mathrm{GHZ})}-S_{_{1}}^{(\mathrm{GHZ})}S_{_{4}}^{(\mathrm{GHZ})}\\
     \end{array}
\end{equation}
Now consider the following density matrices
\begin{equation}\label{pptghz}
    \rho_{_{\pm}}=\frac{1}{16}\big[I_{_{16}}\pm\frac{1}{2}(S_{_{1}}^{(\mathrm{GHZ})}+
  S_{_{1}}^{(\mathrm{GHZ})}S_{_{2}}^{(\mathrm{GHZ})}
    +S_{_{1}}^{(\mathrm{GHZ})}S_{_{4}}^{(\mathrm{GHZ})}-
    S_{_{1}}^{(\mathrm{GHZ})}S_{_{2}}^{(\mathrm{GHZ})}S_{_{4}}^{(\mathrm{GHZ})})\big].
\end{equation}
One can easily check that $ \rho_{_{\pm}}$ are PPT entangled
states and can be detected by the above SEWs, i.e.,
\begin{equation}
Tr(^{i}\mathcal{W}_{_{GHZ}}^{(4)} \rho_{_{-}})=-\frac{1}{2}\qquad
\mathrm{for}\;\;\;i=1,2,
\end{equation}
and
\begin{equation}
Tr(^{i}\mathcal{W}_{_{GHZ}}^{(4)} \rho_{_{+}})=-\frac{1}{2}\qquad
\mathrm{for}\;\;\;i=3,4,5
\end{equation}
which means that all SEWs stated in Eq. (\ref{fouqubitGHZ}) are
nd-SEWs. On the other hand, by the (\ref{optgenform}),
$^{3}\mathcal{W}_{_{GHZ}}^{(4)}$ and
$^{4}\mathcal{W}_{_{GHZ}}^{(4)}$ are optimal SEWs.
\par
Moreover, by the following transformations
\begin{equation}\label{localtrans}
    \begin{array}{c}
  ^{i}\mathcal{W}_{_{GHZ}}^{(4)}\longrightarrow\; ^{i}\mathcal{W}'=U_{local}\ ^{i}\mathcal{W}_{_{GHZ}}^{(4)}U_{local}^{\dag}\\
  \rho_{_{\pm}}\longrightarrow
\rho_{_{\pm}}'=U_{local}\rho_{_{\pm}} U_{local}^\dag \\
\end{array}
\end{equation}
where $U_{local}$ may be  any local unitary Clifford operation we
can get the new nd-SEWs $^{i}\mathcal{W}'$  which can detect the
PPT entangled states $\rho_{_{\pm}}'$. It is necessary to mention
that local unitary operations transform a PPT entangled state to a
PPT one.
\subsection{Decomposability of
$\mathcal{W}_{_{C}}^{(n)}$}\label{clusterdec}
Since  the three-qubit cluster SEWs are transformed to three-qubit
GHZ SEWs  by local unitary Clifford operations as in Eq.
(\ref{localghzcluster}) therefore they are also d-SEWs. For more
than three-qubit the discussion is similar to the GHZ one. The
SEWs
\begin{equation}\label{ineq4}
    I_{_{2^n}}-S_{_{2m-1}}^{(\mathrm{C})}-S_{_{2m}}^{(\mathrm{C})}-
S_{_{2m-1}}^{(\mathrm{C})}S_{_{2m}}^{(\mathrm{C})}
\end{equation}
are optimal d-SEWs since they have positive partial transpose with
respect to the particle $2m-1$ or $2m$. Again a convex cone which
may be formed by connecting every  points of Eq. (\ref{ineq4}) in
the space of parameters to its opposite positive hyper-planes
\begin{equation}\label{}
    a_{_{0}}+\sum_{j=1}^{n'} (-1)^{i_{2j}}
a_{_{2j}}+(-1)^{i_{2m-1}}a_{_{2m-1}}+(-1)^{i_{2m-1}+i_{2m}}a_{_{2m-1,2m}}\geq0
 \end{equation}
for all $(i_{1},i_{2},...,i_{n})\in \{0,1\}^{n}$,  are d-SEWs.
\par
For illustration,  we discuss the odd case of 4-qubit cluster SEW
 in detail. From (\ref{witclod}), we have
$$
\mathcal{W'}_{_{C}}^{(4)}=a_{_{0}}I_{_{2^4}}+\sum_{k=0}^{1}
a_{_{2k+1}}S_{_{2k+1}}^{(\mathrm{C})}
+a_{_{2}}S_{_{2}}^{(\mathrm{C})}+a_{_{2,3}}S_{_{2}}^{(\mathrm{C})}S_{_{3}}^{(\mathrm{C})}
$$
Using the local Clifford operations, all the 14 Hermitian
operators corresponding to boundary half-spaces of the feasible
region can be obtained only from the following 3 ones
\begin{equation}\label{}
    \begin{array}{c}
  ^{1}\mathcal{W'}_{_{C}}^{(4)}=I_{_{2^4}}+S_{_{2}}^{(\mathrm{C})}+S_{_{3}}^{(\mathrm{C})}
  +S_{_{2}}^{(\mathrm{C})}S_{_{3}}^{(\mathrm{C})}, \\
  ^{2}\mathcal{W'}_{_{C}}^{(4)}=I_{_{2^4}}-S_{_{2}}^{(\mathrm{C})}-S_{_{3}}^{(\mathrm{C})}
  -S_{_{2}}^{(\mathrm{C})}S_{_{3}}^{(\mathrm{C})},\\
  ^{3}\mathcal{W'}_{_{C}}^{(4)}=I_{_{2^4}}-S_{_{1}}^{(\mathrm{C})}-S_{_{2}}^{(\mathrm{C})}
  -S_{_{2}}^{(\mathrm{C})}S_{_{3}}^{(\mathrm{C})}. \\
\end{array}
\end{equation}
Among the above operators, $^{1}\mathcal{W'}_{_{C}}^{(4)}$ is
positive  since
$^{1}\mathcal{W'}_{_{C}}^{(4)}=(I+S_{_{2}}^{(\mathrm{C})})(I+S_{_{3}}^{(\mathrm{C})})$,
and if we take partial transpose of the second one with respect to
second particle we get
\begin{equation}\label{}
(^{2}\mathcal{W'}_{_{C}}^{(4)})^{T_{_{2}}}=(I-S_{_{2}}^{(\mathrm{C})})(I-S_{_{3}}^{(\mathrm{C})})\geq0,
\end{equation}
so $^{2}\mathcal{W'}_{_{C}}^{(4)}$ is an optimal d-SEW.
\par
Although we could  not  find  bound entangled states which can be
detected by exactly soluble cluster SEWs however we will be able
to find such entangled states for approximately soluble cluster
SEWs as discussed in section \ref{sectionaprox} and therefore we
postpone  to  subsection \ref{subsecaproxcluster} for more
details.
\section{Separable and Entangled stabilizer states}
Once again consider the general form of operators which is the
same as Eq. (\ref{wgen}), i.e.,
\begin{equation}\label{den1}
 \rho:=\sum_{j_{1},j_{2},...,j_{n-k}=0}^1
b_{j_{1},j_{2},...,j_{n-k}}{S_{_{1}}}^{j_{1}}
{S_{_{2}}}^{j_{2}}...S_{_{n-k}}^{j_{n-k}}=c_{0}I_{_{2^n}}+\sum_{j\neq0}c_{j}
A_{j}
\end{equation}
where for simplicity we have renamed the $S_{_{1}}^{j_{1}}
{S_{_{2}}}^{j_{2}}...S_{_{n-k}}^{j_{n-k}}$ and
$b_{j_{1},j_{2},...,j_{n-k}}$ by $A_{j}$ and $c_{j}$ respectively.
Positivity of $\rho$ together with
$b_{0,0,...,0}=c_{0}=\frac{1}{2^n} $ make (\ref{den1}) a density
matrix. On the other hand, we assert that  the conditions
\begin{equation}\label{pptdens}
\sum_{j_{1},j_{2},...,j_{m}=0}^1 |b_{j_{1},j_{2},...,j_{m}}|\leq
\frac{1}{2^{n-1}}\quad \mathrm{or} \quad   \sum_{j\neq0}
|c_{j}|\leq \frac{1}{2^{n}}
\end{equation}
yields separable state. To see this we note that for any element
$A_{j}$ of $\mathcal{S}_{n-k}$ the operator $I+A_{_{j}}$ is
separable because it is the projection operator on the space
spanned by the pure product eigenvectors of $A_{_{j}}$
corresponding to the eigenvalues $+1$. So any convex combination
of the operators $I+ A_{_{j}}$ such as
\begin{equation}\label{den2}
\varrho_{_{sep}}:=\frac{\mu}{2^n}I_{_{2^n}}
+\frac{(1-\mu)}{2^n}\sum_{j\neq0} p_{j}(I_{_{2^n}}+A_{_{j}})=
\frac{I_{_{2^n}}}{2^n} +\frac{(1-\mu)}{2^n}
\sum_{j\neq0}p_{j}A_{j}
\end{equation}
is separable where $\sum_{j\neq0} p_{j}=1$ and $0\leq\mu\leq1$.
The same statement holds if we replace some $I+ A_{_{j}}$ by $I-
A_{_{j}}$ in  the above equation. Now if we consider all  $c_{j}$
to be positive in Eq. (\ref{den1}) and rename
$\frac{(1-\mu)}{2^n}p_{j}$ by $c_{j}$ (with $j\neq0$) we conclude
that the condition (\ref{pptdens}) is satisfied and therefore  $
\rho$ is separable. For the cases that some $c_{j}$ are negative
it is enough to replace some $I+ A_{_{j}}$ by $I- A_{_{j}}$ in the
Eq. (\ref{den2}) and  proceed the same way as described above.
Consequently we get a family of separable states expressed in
terms of the elements of the stabilizer group provided that the
condition (\ref{pptdens}) satisfies. In the following,  some
entangled states including PPT ones which can be detected by GHZ
and cluster SEWs are introduced.
\subsection{Entangled states which can be detected by $\mathcal{W}_{_{GHZ}}^{(n)}$ }
Now we assert that  GHZ stabilizer EWs can detect some mixed
density matrices. To this aim consider the following operator
\begin{equation}\label{density}
\rho_{_{GHZ}}^{(n)}:=\sum_{j_{1},j_{2},...,j_{n}=0}^1
b_{j_{1},j_{2},...,j_{n}}{S_{_{1}}^{(\mathrm{GHZ})}}^{j_{1}}
{S_{_{2}}^{(\mathrm{GHZ})}}^{j_{2}}...{S_{_{n}}^{(\mathrm{GHZ})}}^{j_{n}}
\end{equation}
which due to tracelessness of $S_{_{i}}^{(\mathrm{GHZ})}$ the
condition $Tr(\rho_{_{GHZ}}^{(n)})=1$ gives
$b_{_{0,0,...,0}}=\frac{1}{2^n}$ and the positivity of density
matrix impose
\begin{equation}\label{posdensity}
\sum_{j_{1},j_{2},...,j_{n}=0}^1 (-1)^{
i_{1}j_{1}+i_{2}j_{2}+...+i_{n}j_{n}}\
b_{j_{1},j_{2},...,j_{n}}\geq0\quad ,\quad \forall \
(i_{1},i_{2},...,i_{n})\in \{0,1\}^{n}
\end{equation}
to its eigenvalues. An interesting case is when all coefficients
are equal to $ b_{j_{1},j_{2},...,j_{n}}=\frac{1}{2^n}$ which is
coincides with the n-qubit GHZ state
$$
|\psi_{_{00...0}}\rangle\langle
\psi_{_{00...0}}|=\frac{1}{2^n}\prod_{j=1}^n
(I+S_{_{j}}^{(\mathrm{GHZ})}).
$$
This density matrix has $2^n$ terms which except
$b_{_{0,0,...,0}}$ the other are arbitrary parameters with the
constraints in Eq. (\ref{posdensity}). These $2^n$ constraints
forms a simplex polygon in a $2^n-1$ dimensional space with
coordinate variables $b_{j_{1},j_{2},...,j_{n}}$ (excepted
$b_{_{0,0,...,0}}$). Furthermore if we want $\rho_{_{GHZ}}^{(n)}$
becomes a PPT entangled state in the sense that its partial
transpose is positive definite with respect to any particle, i.e.,
${\rho_{_{GHZ}}^{(n)}}^{T_{i}} \geq0$ with $i=1,...,n$ then we
must have
$$
 \sum_{j_{1},j_{2},...,j_{n}=0}^1\left\{
(-1)^{ i_{1}}\ b_{1,j_{2},...,j_{n}}+ (-1)^{
i_{2}j_{2}+i_{3}j_{3}+...+i_{n}j_{n}}\
b_{0,j_{2},...,j_{n}}\right\}\geq0\quad , \forall \
(i_{1},i_{2},...,i_{n})\in \{0,1\}^{n}
$$
Introducing the new parameters  $
b_{_{i}}=b_{_{0,...,0,1,0,...,0}}$ with a $1$ in the $i$th
position, and $b_{_{1,j}}=b_{_{1,0,...,0,1,0,...,0}}$ with a $1$
in the $j$th position , and using the orthogonality
(\ref{staborthogonal}) of $S_{_{i}}$'s, then the condition for
detectability of $\rho_{_{GHZ}}^{(n)}$ by
$\mathcal{W}_{_{GHZ}}^{(n)}$ can be written as
$$
Tr(\mathcal{W}_{_{GHZ}}^{(n)}
\rho_{_{GHZ}}^{(n)})=\frac{a_{_{0}}}{2^n}+ \sum_{k=1}^n
a_{_{k}}b_{_{k}}+\sum_{k=1}^{n'} a_{_{1,2k}}b_{_{1,2k}}<0
$$
\subsection{Entangled states which can be detected by $\mathcal{W}_{_{C}}^{(n)}$ }
Now we assert that the above cluster stabilizer EWs can detect
some mixed density matrices. To this aim consider the following
operator
\begin{equation}\label{density}
\rho_{_{C}}^{(n)}:=\sum_{j_{1},j_{2},...,j_{n}=0}^1
b_{j_{1},j_{2},...,j_{n}}{S_{_{1}}^{(\mathrm{C})}}^{j_{1}}{S_{_{2}}^{(\mathrm{C})}}^{j_{2}}...{S_{_{n}}^{(\mathrm{C})}}^{j_{n}}
\end{equation}
which due to traceless of $S_{_{i}}^{(\mathrm{C})}$ the condition
$Tr(\rho)=1$ gives  $b_{_{0,0,...,0}}=\frac{1}{2^n}$ and the
positivity of density matrix impose
\begin{equation}\label{posdensity}
\sum_{j_{1},j_{2},...,j_{n}=0}^1 (-1)^{
i_{1}j_{1}+i_{2}j_{2}+...+i_{n}j_{n}}\
b_{j_{1},j_{2},...,j_{n}}\geq0\quad ,\quad \forall \
(i_{1},i_{2},...,i_{n})\in \{0,1\}^{n}
\end{equation}
to the its eigenvalues. An interesting case is when all coefficients
are equal to $ b_{j_{1},j_{2},...,j_{n}}=\frac{1}{2^n}$ which is
coincides with the n-qubit cluster state
$$
|\mathrm{C}\rangle\langle \mathrm{C}|=\frac{1}{2^n}\prod_{j=1}^n
(I+S_{_{j}}^{(\mathrm{C})}).
$$
In order to $\rho_{_{C}}^{(n)}$ can be detected by an odd case
$\mathcal{W'}_{_{C}}^{(n)}$, we must have
$$
Tr(\mathcal{W'}_{_{C}}^{(n)}
\rho_{_{C}}^{(n)})=\frac{a_{_{0}}}{2^n}+ \sum_{k=1}^n
a_{_{2k+1}}b_{_{2k+1}}+a_{_{2m}}b_{_{2m}}+a_{_{2m,2m+1}}b_{_{2m,2m+1}}<0.
$$
\section{Approximate stabilizer EWs}\label{sectionaprox}
So far, we have considered SEWs which can be exactly solved by LP
method. In this section, we consider approximately soluble SEWs
which come from by adding some other members of stabilizer group
to exactly soluble SEWs. In all of the SEWs discussed in section
\ref{sec1}, the boundary half-spaces arise from the vertices which
themselves come from pure product states and the resulting
inequalities did not offend against the convex hull of vertices at
all. But  by adding some terms to exactly soluble SEWs, it may be
happen that the feasible region be convex with curvature on some
boundaries and the problem can not be solved by exactly LP method.
In these cases the linear constraints no longer arise from convex
hull of the vertices coming from pure product states. Hence we
transform such problem to approximately soluble LP one. Our
approach is to draw the hyper-planes tangent to feasible region
and parallel to hyper-planes coming from vertices and in this way
we enclose the feasible regions  by such hyper-planes. It is clear
that in this extension, the vertices no longer arise from pure
product states.
\subsection{Approximate n-qubit GHZ SEWs}
For the even case of $\mathrm{GHZ}$ SEWs we add one of the
statements $S_{_{1}}^{(\mathrm{GHZ})}S_{_{2l+1}}^{(\mathrm{GHZ})}$
($l=1,...,n''$) to Eq. (\ref{witghzev}) as
\begin{equation}\label{}
    \mathcal{W}_{_{GHZ_{(ap)}}}^{(n)}=a_{_{0}}I_{_{2^n}}+\sum_{k=1}^n
a_{_{k}}S_{_{k}}^{(\mathrm{GHZ})}+\sum_{k=1}^{n'}a_{_{1,2k}}S_{_{1}}^{(\mathrm{GHZ})}S_{_{2k}}^{(\mathrm{GHZ})}
+a_{_{1,2l+1}}S_{_{1}}^{(\mathrm{GHZ})}S_{_{2l+1}}^{(\mathrm{GHZ})}
\end{equation}
and  try to solve it by the LP method.  The eigenvalues of
$\mathcal{W}_{_{GHZ_{(ap)}}}^{(n)}$ are
$$
a_{_{0}}+\sum_{j=1}^n (-1)^{i_{j}}
a_{_{j}}+\sum_{k=1}^{n'}(-1)^{i_{1}+
i_{2k}}a_{_{1,2k}}+(-1)^{i_{1}+ i_{2l+1}}a_{_{1,2l+1}}\quad ,\quad
\forall \ (i_{1},i_{2},...,i_{n})\in \{0,1\}^{n}
$$
The coordinates of the vertices which  arise from pure product
vectors are listed in the table 3
\begin{table}[h]
\renewcommand{\arraystretch}{1}
\addtolength{\arraycolsep}{-2pt}
$$
\begin{array}{|c|c|}\hline
\mathrm{Product \ state} &
(P_{2},P_{3},...,P_{n-1},P_{n},P_{1},P_{1,2},P_{1,4},...,P_{1,2n'-2},P_{1,2n'},P_{1,2l+1})
\\ \hline
  |\Psi^{\pm}\rangle & (0,0,...,0,0,\pm1,0,0,...,0,0,0) \\
  \Lambda_{_{1}} |\Psi^{\pm}\rangle & (0,0,...,0,0,0,\pm1,0,...,0,0,0) \\
  \vdots & \vdots \\
  \Lambda_{_{n'}} |\Psi^{\pm}\rangle & (0,0,...,0,0,0,0,0,...,0,\pm1,0) \\
  \Lambda_{_{2l+1}}^{(\mathrm{ap})} |\Psi^{\pm}\rangle & (0,0,...,0,0,0,0,0,...,0,0,\pm1) \\
  \hline
  \Xi_{i_{2},...,i_{n}}|\Psi^{+}\rangle & \left((-1)^{i_{_{2}}},
  (-1)^{i_{_{2}}+i_{_{3}}},...,(-1)^{i_{_{n-2}}+i_{_{n-1}}},(-1)^{i_{_{n-1}}+i_{_{n}}},0,0,0,...,0,0,0\right)
    \\ \hline
\end{array}
$$
\caption{\small{The product vectors which seem to be the vertices
 for $\mathcal{W}_{_{GHZ_{(ap)}}}^{(n)}$.}}
\renewcommand{\arraystretch}{1}
\addtolength{\arraycolsep}{-3pt}
\end{table}

where
$$
\Lambda_{_{2l+1}}^{(\mathrm{ap})}=\left({M^{(2l)}}\right)^{\dagger}
M^{(2l+1)}.
$$
Choosing any $N_{1}=n+n'+1$ points among $N_{2}=2^{n-1}+n'+2$
above points  we get the following $\mathbf{C}_{N_{1}}^{N_{2}}$
half-spaces
\begin{equation}\label{ineapprev}
 |P_{_{1}}\pm P_{_{j}}+\sum_{k=1}^{n'}(-1)^{i_{k}}P_{_{1,2k}}+(-1)^{i_{n'+1}}P_{_{1,2l+1}}|\leq \mu_{max}=?,\quad\quad
  j=2,...,n,
\end{equation}
where $ (i_{1},...,i_{n'+1})\in \{0,1\}^{n'+1}.$ But calculations
show that the inequalities offend against 1 up to
$\mu_{max}=\frac{1+\sqrt{2}}{2}$ (see appendix II). This shows
that the problem does not lie in the realm of exactly soluble LP
problems and we have to use approximate LP. To this aim, we shift
aforementioned  hyper-planes parallel to themselves such that they
reach to maximum value $\mu_{max}=\frac{1+\sqrt{2}}{2}$. On the
other hand the maximum shifting is where the  hyper-planes become
tangent to convex region coming from pure product states and in
this manner we will be able to encircle the  feasible region by
the half-spaces
\begin{equation}\label{}
    \begin{array}{c}
       |P_{_{1}}+ P_{_{2j}}+\sum_{k=1}^{n'}P_{_{1,2k}}+P_{_{1,2l+1}}|\leq
\frac{1+\sqrt{2}}{2}\quad\quad j=1,...,n^{'} \\
       |P_{_{1}}+
P_{_{2l+1}}+\sum_{k=1}^{n'}P_{_{1,2k}}+P_{_{1,2l+1}}|\leq
\frac{1+\sqrt{2}}{2} \\
       P_{_{1}}+ P_{_{2j+1}}+\sum_{k=1}^{n'}P_{_{1,2k}}\leq
\frac{1+\sqrt{2}}{2}\quad\quad l\neq j=1,...,n^{''} \\
        P_{_{1}}\leq1 \\
     \end{array}
\end{equation}
where  again we have used the Clifford group and write just the
generating half-spaces. Due to the above inequalities the problem
is reduced to the LP problem
$$
\begin{array}{c}
  \hspace{-4cm}\mathrm{minimize} \quad\quad
Tr(\mathcal{W}_{_{GHZ_{(ap)}}}^{(n)}|\gamma\rangle\langle\gamma|) \\
  \mathrm{s.t.}\quad
\left\{\begin{array}{c}
  |P_{_{1}}\pm
P_{_{j}}+\sum_{k=1}^{n'}(-1)^{i_{k}}P_{_{1,2k}}+(-1)^{i_{n'+1}}P_{_{1,2l+1}}|\leq
\frac{1+\sqrt{2}}{2} \quad j=2,...,n\\
\hspace{-5.7cm} P_{_{i}}\leq1 \quad \;\;\quad \;\; i=1,...,n\\
\hspace{-5.5cm} P_{_{1,2k}}\leq1 \quad\quad  k=1,...,n'\\
\hspace{-8.1cm}  P_{_{1,2l+1}}\leq1 \\
\end{array}
\right.
\end{array}
$$
for all $ (i_{1},i_{2},...,i_{n'},i_{n'+1})\in \{0,1\}^{n'+1}$,
where it can be solved by simplex method. \\
The intersections of the half-spaces in the above equation form a
convex polygon whose  vertices lie at  any permutation
$P'_{_{1}},P'_{_{j}}$, $P'_{_{1,2k}}(k=1,...,n')$ and
$P'_{_{1,2l+1}}$ with a given $j\; (j=2,...,n)$ of the points
listed in table 4
\begin{table}[h]
\renewcommand{\arraystretch}{1}
\addtolength{\arraycolsep}{-2pt}
$$
\begin{array}{|c|}\hline
(P'_{2},P'_{3},...,P'_{n-1},P'_{n},P'_{1},P'_{1,2},P'_{1,4},...,P'_{1,2n'-2},P'_{1,2n'},P'_{1,2l+1})
\\ \hline
   \big((-1)^{i_{2}},(-1)^{i_{3}},...,(-1)^{i_{n-1}},(-1)^{i_{n}},(-1)^{i_{1}},(-1)^{i_{1,2}},(-1)^{i_{1,4}},...,(-1)^{i_{1,2n'-2}},(-1)^{i_{1,2n'}},\frac{\sqrt{2}-3}{2}\big) \\
 \ni P'_{_{1}}+
  P'_{_{j}}+\sum_{k=1}^{n'}P'_{_{1,2k}}=2\\
\hline
    \big((-1)^{i_{2}},(-1)^{i_{3}},...,(-1)^{i_{n-1}},(-1)^{i_{n}},(-1)^{i_{1}},(-1)^{i_{1,2}},(-1)^{i_{1,4}},...,(-1)^{i_{1,2n'-2}},(-1)^{i_{1,2n'}},\frac{3-\sqrt{2}}{2}\big) \\
  \ni P'_{_{1}}+
  P'_{_{j}}+\sum_{k=1}^{n'}P'_{_{1,2k}}=-2 \\\hline
  \big((-1)^{i_{2}},(-1)^{i_{3}},...,(-1)^{i_{n-1}},(-1)^{i_{n}},(-1)^{i_{1}},(-1)^{i_{1,2}},(-1)^{i_{1,4}},...,(-1)^{i_{1,2n'-2}},(-1)^{i_{1,2n'}},\frac{\sqrt{2}-1}{2}\big)
  \\
\ni  P'_{_{1}}+
  P'_{_{j}}+\sum_{k=1}^{n'}P'_{_{1,2k}}=1\\ \hline
  \big((-1)^{i_{2}},(-1)^{i_{3}},...,(-1)^{i_{n-1}},(-1)^{i_{n}},(-1)^{i_{1}},(-1)^{i_{1,2}},(-1)^{i_{1,4}},...,(-1)^{i_{1,2n'-2}},(-1)^{i_{1,2n'}},\frac{1-\sqrt{2}}{2}\big)
  \\
\ni P'_{_{1}}+
  P'_{_{j}}+\sum_{k=1}^{n'}P'_{_{1,2k}}=-1 \\ \hline
  \end{array}
$$
\caption{\small{The  coordinates of vertices for
 $\mathcal{W}_{_{GHZ_{(ap)}}}^{(n)}$.}}
\renewcommand{\arraystretch}{1}
\addtolength{\arraycolsep}{-3pt}
\end{table}
where $P'$ 's are defined by shifting the $P$ 's  for all
$(i_{1},...,i_{n},i_{1,2},...,i_{1,2n'})\in
{\{0,1\}}^{n+n'}$.\\
So in order that the expectation value of
$\mathcal{W}_{_{GHZ_{(ap)}}}^{(n)}$ be non-negative over any pure
product state, the following inequalities and any inequality
obtained from them by permuting the parameters
$a_{_{1}},a_{_{j}}$, $a_{_{1,2k}}(k=1,...,n')$ and $a_{_{1,2l+1}}$
with a given $j$ for $j=2,...,n$, must be fulfilled
$$
\begin{array}{c}
  a_{_{0}}+\sum_{k=1}^{n}(-1)^{i_{k}}a_{_{k}}+\sum_{k=1}^{n'}(-1)^{i_{1,2k}}a_{_{1,2k}}+
\frac{\sqrt{2}-3}{2}a_{_{1,2l+1}}\geq 0  \\
\mathrm{such}\;\mathrm{that}\;\;(-1)^{i_{1}}+(-1)^{i_{j}}+\sum_{k=1}^{n'}(-1)^{i_{1,2k}}=2 \\
 \end{array}
$$
$$
\begin{array}{c}
  a_{_{0}}+\sum_{k=1}^{n}(-1)^{i_{k}}a_{_{k}}+\sum_{k=1}^{n'}(-1)^{i_{1,2k}}a_{_{1,2k}}+
\frac{3-\sqrt{2}}{2}a_{_{1,2l+1}}\geq 0  \\
\mathrm{such}\;\mathrm{that}\;\;(-1)^{i_{1}}+(-1)^{i_{j}}+\sum_{k=1}^{n'}(-1)^{i_{1,2k}}=-2 \\
 \end{array}
$$
$$
\begin{array}{c}
  a_{_{0}}+\sum_{k=1}^{n}(-1)^{i_{k}}a_{_{k}}+\sum_{k=1}^{n'}(-1)^{i_{1,2k}}a_{_{1,2k}}+
\frac{\sqrt{2}-1}{2}a_{_{1,2l+1}}\geq 0  \\
\mathrm{such}\;\mathrm{that}\;\;(-1)^{i_{1}}+(-1)^{i_{j}}+\sum_{k=1}^{n'}(-1)^{i_{1,2k}}=1 \\
 \end{array}
$$
\begin{equation}\label{}
\begin{array}{c}
  a_{_{0}}+\sum_{k=1}^{n}(-1)^{i_{k}}a_{_{k}}+\sum_{k=1}^{n'}(-1)^{i_{1,2k}}a_{_{1,2k}}+
\frac{1-\sqrt{2}}{2}a_{_{1,2l+1}}\geq 0  \\
\mathrm{such}\;\mathrm{that}\;\;(-1)^{i_{1}}+(-1)^{i_{j}}+\sum_{k=1}^{n'}(-1)^{i_{1,2k}}=-1 \\
 \end{array}
\end{equation}
Similarly, one could repeat this approximation for the odd case of
GHZ SEWs like above.
\par
As the case of exactly soluble GHZ SEWs, we assert that there
exist some nd-SEWs among the approximately soluble  GHZ SEWs. To
see this, consider the four-qubit GHZ SEWs
\begin{equation}\label{}
    \mathcal{W}_{\pm}=\frac{1+\sqrt{2}}{2}\ I_{_{16}}+S_{_{1}}^{(\mathrm{GHZ})}+S_{_{2}}^{(\mathrm{GHZ})}+
  S_{_{1}}^{(\mathrm{GHZ})}S_{_{2}}^{(\mathrm{GHZ})}+S_{_{1}}^{(\mathrm{GHZ})}S_{_{4}}^{(\mathrm{GHZ})}
  \pm S_{_{1}}^{(\mathrm{GHZ})}S_{_{3}}^{(\mathrm{GHZ})}
\end{equation}
which both can detect the PPT entangled state in Eq.
(\ref{pptghz}) with
$Tr(\mathcal{W}_{\pm}\rho_{_{-}})=-\frac{2-\sqrt{2}}{2}\simeq
-0.29$.
\subsection{Approximated n-qubit Cluster
SEWs}\label{subsecaproxcluster} For the odd case of cluster SEWs
we add one of the statements
$S_{_{2m-1}}^{(\mathrm{C})}S_{_{2m}}^{(\mathrm{C})}$ to Eq.
(\ref{witclod}) as
\begin{equation}\label{}
    \mathcal{W'}_{_{C_{(ap.)}}}^{(n)}=a_{_{0}}I_{_{2^n}}+\sum_{k=0}^{n''}
a_{_{2k+1}}S_{_{2k+1}}^{(\mathrm{C})}
+a_{_{2m}}S_{_{2m}}^{(\mathrm{C})}+a_{_{2m,2m+1}}S_{_{2m}}^{(\mathrm{C})}S_{_{2m+1}}^{(\mathrm{C})}+
a_{_{2m-1,2m}}S_{_{2m-1}}^{(\mathrm{C})}S_{_{2m}}^{(\mathrm{C})},
\end{equation}
where $m=1,...,n''$. The eigenvalues of
$\mathcal{W'}_{_{C_{(ap.)}}}^{(n)}$ for all
$(i_{1},i_{2},...,i_{n})\in \{0,1\}^{n}$ are
$$
a_{_{0}}+\sum_{j=0}^{n''} (-1)^{i_{2j+1}}
a_{_{2j+1}}+(-1)^{i_{2m}}a_{_{2m}}+(-1)^{i_{2m}+i_{2m+1}}a_{_{2m,2m+1}}+(-1)^{i_{2m-1}+i_{2m}}a_{_{2m-1,2m}}
$$
The coordinates of the vertices which  arise from pure product
vectors are listed in the Table 5,
\begin{table}[h]
\renewcommand{\arraystretch}{1.2}
\addtolength{\arraycolsep}{-2pt}
$$
\begin{array}{|c|c|}\hline
 \mathrm{Product\ state} & (P_{1},P_{3},...,P_{2m-3},P_{2m-1},P_{2m},P_{2m+1},P_{2m+3},..., P_{2n''+1},P_{2m,2m+1}
 ,P_{2m-1,2m}) \\ \hline
  \Lambda_{_{i_{1},i_{2},...,i_{n''+1}}}^{(odd)}|\Phi\rangle & \big((-1)^{i_{1}},(-1)^{i_{2}},...,
  (-1)^{i_{m-2}},(-1)^{i_{m-1}},0,(-1)^{i_{m}},(-1)^{i_{m+1}},...,(-1)^{i_{n''+1}},0,0\big) \\
   \hline
  {\Lambda'}_{_{i_{1},i_{2},...,i_{n''+1}}}^{(odd)}|\Phi\rangle & \big((-1)^{i_{1}},(-1)^{i_{2}},...,
  (-1)^{i_{m-2}},0,\pm1,0,(-1)^{i_{m+1}},...,(-1)^{i_{n''+1}},0,0\big)\\
  \hline
  {\Lambda''}_{_{i_{1},i_{2},...,i_{n''+1}}}^{(odd)}|\Phi\rangle & ((-1)^{i_{1}},(-1)^{i_{2}},...
  ,(-1)^{i_{m-2}},0,0,0,(-1)
  ^{i_{m+1}},...,(-1)^{i_{n''+1}},(-1)^{i_{m}},0)\\
  \hline
  {\Lambda'''}_{_{i_{1},i_{2},...,i_{n''+1}}}^{(odd)}|\Phi\rangle & ((-1)^{i_{1}},(-1)^{i_{2}},...
  ,(-1)^{i_{m-2}},0,0,0,(-1)
  ^{i_{m+1}},...,(-1)^{i_{n''+1}},0,(-1)^{i_{m}})\\
  \hline
\end{array}
$$
\caption{\small{The product vectors which seem to be the vertices
for $\mathcal{W'}_{_{C_{(ap)}}}^{(n)}$.}}
\renewcommand{\arraystretch}{1}
\addtolength{\arraycolsep}{3pt}
\end{table}
where
$$
{\Lambda'''}_{_{i_{1},i_{2},...,i_{n''+1}}}^{(odd)}=\Lambda_{_{i_{1},i_{2},...,i_{n''+1}}}^{(odd)}
  H^{(2m-2)}M^{(2m-1)}H^{(2m-1)}M^{(2m)}
$$
By choosing any $n''+4$ from $2^{n''+1}+3\times2^{n''}$ points,
the following half-spaces achieves
\begin{equation}\label{ineclusterapp}
\begin{array}{c}
  |P_{2m}+(-1)^{i_{1}}P_{2m-1}+(-1)^{i_{2}}P_{2m,2m+1}+(-1)^{i_{3}}P_{2m-1,2m}|\leq\frac{2}{\sqrt{3}}\\
  \hspace{-4.5cm}|P_{2k+1}|\leq1 \quad\quad , \quad\quad    m-1\neq k=0,...,n'' \\
\end{array}
\end{equation}
for all $(i_{1},i_{2},i_{3})\in \{0,1\}^{n}$ (see appendix II).
This shows that the problem does not lie in the realm of exact LP
problems and we have to use approximate LP one. To do so, we shift
aforementioned  hyper-planes parallel to themselves such that they
reach to maximum value $\eta=\frac{2}{\sqrt{3}}$. On the other
hand the maximum shifting is where the  hyper-planes become
tangent to convex region coming from pure product states and in
this manner we will be able to encircle the  feasible region by
the half-spaces
\begin{equation}\label{appcl}
    \begin{array}{c}
      |P_{2m}+P_{2m-1}+P_{2m,2m+1}+P_{2m-1,2m}|\leq
      \frac{2}{\sqrt{3}}\\
      P_{1}\leq1
    \end{array}
\end{equation}
where  again we have used the Clifford group and write just the
generating half-spaces. Due to the above inequalities the problem
is approximately reduced to
\begin{equation}\label{}
    \begin{array}{c}
     \hspace{-8cm}\mathrm{minimize} \quad
     Tr(\mathcal{W'}_{_{C_{(ap.)}}}^{(n)}|\gamma\rangle\langle\gamma|) \\
     \mathrm{s.t.}\quad \left\{
      \begin{array}{c}
     |P_{2m}+(-1)^{i_{1}}P_{2m-1}+(-1)^{i_{2}}P_{2m,2m+1}+(-1)^{i_{3}}P_{2m-1,2m}|\leq\frac{2}{\sqrt{3}}\quad,
     \forall \ (i_{1},i_{2},i_{3})\in \{0,1\}^{n}\\
     |P_{2m}+(-1)^{i_{1}}P_{2m+1}+(-1)^{i_{2}}P_{2m,2m+1}+(-1)^{i_{3}}P_{2m-1,2m}|\leq\frac{2}{\sqrt{3}}\quad,
     \forall \ (i_{1},i_{2},i_{3})\in \{0,1\}^{n}\\
     \hspace{-10cm}|P_{2k+1}|\leq1 \quad , \quad    k=0,...,n''\\
     \hspace{-13.4cm} |P_{2m}|\leq1\\
     \hspace{-12.5cm} |P_{2m,2m+1}|\leq1\\
     \hspace{-12.5cm} |P_{2m-1,2m}|\leq1\\
      \end{array}
     \right.
    \end{array}
\end{equation}
which can be solved by LP method. \\
The intersections of the half-spaces in the above equation form a
convex polygon whose  vertices lie at any permutation of the
coordinates $P_{2m-1},P_{2m},P_{2m,2m+1},P_{2m-1,2m}$ of the
points listed in Table 12
\begin{table}[h]
\renewcommand{\arraystretch}{1}
\addtolength{\arraycolsep}{-2pt}
$$
\small{
\begin{array}{|c|}\hline
(P'_{1},P'_{3},...,P'_{2m-3},P'_{2m-1},P'_{2m},P'_{2m+1},P'_{2m+3},...,
P'_{2n''+1},P'_{2m,2m+1}
 ,P'_{2m-1,2m})\\ \hline
   \big((-1)^{i_{1}},(-1)^{i_{3}},...,
  (-1)^{i_{2m-3}},(-1)^{i_{2m-1}},(-1)^{i_{2m}},(-1)^{i_{2m+1}},(-1)^{i_{2m+3}},...,(-1)^{i_{2n''+1}},(-1)^{i_{2m,2m+1}},\frac{2-\sqrt{3}}{\sqrt{3}}\big) \\
  \ni P'_{2m}+P'_{2m-1}+P'_{2m,2m+1}+P'_{2m-1,2m}=1\\
\hline
   \big((-1)^{i_{1}},(-1)^{i_{3}},...,
  (-1)^{i_{2m-3}},(-1)^{i_{2m-1}},(-1)^{i_{2m}},(-1)^{i_{2m+1}},(-1)^{i_{2m+3}},...,(-1)^{i_{2n''+1}},(-1)^{i_{2m,2m+1}},\frac{\sqrt{3}-2}{\sqrt{3}}\big) \\
   \ni P'_{2m}+P'_{2m-1}+P'_{2m,2m+1}+P'_{2m-1,2m}=-1\\\hline
\end{array}}
$$
\caption{\small{The  coordinates of vertices for
$\mathcal{W'}_{_{C_{(ap)}}}^{(n)}$.}}
\renewcommand{\arraystretch}{1}
\addtolength{\arraycolsep}{-3pt}
\end{table}
and any permutation of the coordinates
$P_{2m},P_{2m+1},P_{2m,2m+1},P_{2m-1,2m}$ of the points listed in
Table 13
\begin{table}[h]
\renewcommand{\arraystretch}{1}
\addtolength{\arraycolsep}{-2pt}
$$
\small{
\begin{array}{|c|}\hline
(P'_{1},P'_{3},...,P'_{2m-3},P'_{2m-1},P'_{2m},P'_{2m+1},P'_{2m+3},...,
P'_{2n''+1},P'_{2m,2m+1}
 ,P'_{2m-1,2m})\\ \hline
  \big((-1)^{i_{1}},(-1)^{i_{3}},...,
  (-1)^{i_{2m-3}},(-1)^{i_{2m-1}},(-1)^{i_{2m}},(-1)^{i_{2m+1}},(-1)^{i_{2m+3}},...,(-1)^{i_{2n''+1}},(-1)^{i_{2m,2m+1}},\frac{2-\sqrt{3}}{\sqrt{3}}\big) \\
  \ni P'_{2m}+P'_{2m+1}+P'_{2m,2m+1}+P'_{2m-1,2m}=1\\
\hline
   \big((-1)^{i_{1}},(-1)^{i_{3}},...,
  (-1)^{i_{2m-3}},(-1)^{i_{2m-1}},(-1)^{i_{2m}},(-1)^{i_{2m+1}},(-1)^{i_{2m+3}},...,(-1)^{i_{2n''+1}},(-1)^{i_{2m,2m+1}},\frac{\sqrt{3}-2}{\sqrt{3}}\big) \\
   \ni P'_{2m}+P'_{2m+1}+P'_{2m,2m+1}+P'_{2m-1,2m}=-1\\\hline
  \end{array}}
$$
\caption{\small{The  vertex  points of approximated FR of
 $\mathcal{W'}_{_{C_{(ap)}}}^{(n)}$.}}
\renewcommand{\arraystretch}{1}
\addtolength{\arraycolsep}{-3pt}
\end{table}
where $P'$ 's are defined by shifting the $P$ 's and for all $(i_{1},...,i_{n},i_{2m,2m+1})\in {\{0,1\}}^{n+1}$.\\
Therefore, to be guaranteed the non-negativity of the expectation
value of $\mathcal{W'}_{_{C_{(ap)}}}^{(n)}$ over all pure product
states, the conditions
$$
\begin{array}{c}
a_{_{0}}+\sum_{k=0}^{n''}(-1)^{i_{2k+1}}a_{_{2k+1}}+(-1)^{i_{2m}}a_{_{2m}}+(-1)^{i_{2m,2m+1}}a_{_{2m,2m+1}}+
\frac{2-\sqrt{3}}{\sqrt{3}}a_{_{2m-1,2m}}\geq 0  \\
\mathrm{such}\;\mathrm{that}\;\;(-1)^{i_{2m-1}}+(-1)^{i_{2m}}+(-1)^{i_{2m-1,2m}}+(-1)^{i_{2m,2m+1}}=1 \\
\end{array}
$$
\begin{equation}\label{}
\begin{array}{c}
a_{_{0}}+\sum_{k=0}^{n''}(-1)^{i_{2k+1}}a_{_{2k+1}}+(-1)^{i_{2m}}a_{_{2m}}+(-1)^{i_{2m,2m+1}}a_{_{2m,2m+1}}+
\frac{\sqrt{3}-2}{\sqrt{3}}a_{_{2m-1,2m}}\geq 0  \\
\mathrm{such}\;\mathrm{that}\;\;(-1)^{i_{2m-1}}+(-1)^{i_{2m}}+(-1)^{i_{2m-1,2m}}+(-1)^{i_{2m,2m+1}}=-1 \\
\end{array}
\end{equation}
and any permutation of parameters $a_{2m-1},a_{2m},a_{2m,2m+1}$ and
$a_{2m-1,2m}$ together with the following conditions
$$
\begin{array}{c}
a_{_{0}}+\sum_{k=0}^{n''}(-1)^{i_{2k+1}}a_{_{2k+1}}+(-1)^{i_{2m}}a_{_{2m}}+(-1)^{i_{2m,2m+1}}a_{_{2m,2m+1}}+
\frac{2-\sqrt{3}}{\sqrt{3}}a_{_{2m-1,2m}}\geq 0  \\
\mathrm{such}\;\mathrm{that}\;\;(-1)^{i_{2m-1}}+(-1)^{i_{2m}}+(-1)^{i_{2m-1,2m}}+(-1)^{i_{2m,2m+1}}=1 \\
\end{array}
$$
\begin{equation}\label{}
\begin{array}{c}
a_{_{0}}+\sum_{k=0}^{n''}(-1)^{i_{2k+1}}a_{_{2k+1}}+(-1)^{i_{2m}}a_{_{2m}}+(-1)^{i_{2m,2m+1}}a_{_{2m,2m+1}}+
\frac{\sqrt{3}-2}{\sqrt{3}}a_{_{2m-1,2m}}\geq 0  \\
\mathrm{such}\;\mathrm{that}\;\;(-1)^{i_{2m-1}}+(-1)^{i_{2m}}+(-1)^{i_{2m-1,2m}}+(-1)^{i_{2m,2m+1}}=-1 \\
\end{array}
\end{equation}
and any permutation of parameters $a_{2m},a_{2m+1},a_{2m,2m+1}$
and $a_{2m-1,2m}$ must be fulfilled. Similarly, one could repeat
this approximate solve for the even case of cluster SEWs just like
above.
\par
Now we discuss the non-decomposability of cluster SEWs mentioned
at the end of subsection \ref{clusterdec}. For more than
three-qubit, the SEWs corresponding to half-spaces (\ref{appcl})
contain some nd-SEWs. As an instance, consider the SEW
\begin{equation}\label{}
  \mathcal{W}=\frac{2}{\sqrt{3}}\;I_{_{16}}-S_{_{1}}^{(C)}-S_{_{2}}^{(C)}-
  S_{_{1}}^{(C)}S_{_{2}}^{(C)}-S_{_{2}}^{(C)}S_{_{3}}^{(C)}
\end{equation}
The expectation value of $\mathcal{W}$ with the following density
matrix
$$
\rho=\frac{1}{16}\big[I_{_{16}}+\frac{1}{2}(S_{_{2}}^{(C)}+
  S_{_{1}}^{(C)}S_{_{2}}^{(C)}
  +S_{_{2}}^{(C)}S_{_{3}}^{(C)}-
  S_{_{1}}^{(C)}S_{_{2}}^{(C)}S_{_{3}}^{(C)})\big]
$$
is $Tr(\mathcal{W} \rho)=-0.345$ which means that $\mathcal{W}$
can detect $\rho$. On the other hand one can easily check that
$\rho$ is a bound entangled state and hence $\mathcal{W}$ is a
nd-SEW.
\section{Conclusion}
We have considered the construction of EWs by using the stabilizer
operators of some given multi-qubit states. It was shown that when
the feasible region is a polygon or can be approximated by a
polygon, the problem is reduced to a LP one. For illustrating the
method, several examples including GHZ, cluster, and exceptional
states EWs have studied in details. The optimality and
decomposability or non-decomposability of SEWs corresponding to
boundary half-planes surrounding the feasible region have examined
and it was shown that the optimality has a close connection with
the common eigenvectors of SEWs. In each instance, it was shown
that the feasible region is a polygon and the Hermitian operators
corresponding to half-planes surrounding it are SEWs or positive.
Also we have showed that, by using the Clifford group operations
one can find vertex points and surrounding half-planes of feasible
region only from a few ones.

\vspace{1cm} \setcounter{section}{0}
 \setcounter{equation}{0}
 \renewcommand{\theequation}{I-\arabic{equation}}
\newpage
{\Large{Appendix I:}}\\
\textbf{Stabilizer theory}\\
 Here we summarize the stabilizer
formalism  and its application to construct an interesting class
of EWs so-called
\emph{stabilizer entanglement witnesses} (SEWs) \cite{guhne1}.\\
The $l=2^k$  (where $ k=0,1,...,n$) stabilizer states
$\{|\psi_{1}\rangle,...,|\psi_{l}\rangle\}$ of $n$ qubits can be
thought of as  representation of an abelian stabilizer group
$\mathcal{S}_{n-k}$ generated by $n-k$ pairwise commuting
Hermitian operators in the Pauli group $\mathcal{G}_{n}$, which
consists of tensor products of the identity $I_{2}$ and the usual
Pauli matrices $\sigma_{x},\sigma_{y}$ and $\sigma_{z}$ together
with an overall phase $\pm1$ or $\pm i$
\cite{preskill,gott1,gott2}.
 The group $\mathcal{S}_{n-k}$ has
$2^{n-k}$ elements where among them we can choose
$S_{_{1}},...,S_{_{n-k}}$ as generators. This group leaves
invariant any state in the stabilizer Hilbert space
$\mathcal{H}_{S}$ spanned by
$\{|\psi_{_{1}}\rangle,...,|\psi_{_{l}}\rangle\}$, i.e.,
\begin{equation}
S|\psi\rangle=|\psi\rangle\quad, \forall \;\; S\in
\mathcal{S}_{n-k} \quad,\; \forall
\;|\psi\rangle\in\mathcal{H}_{S}.
\end{equation}
Similar to the Pauli matrices, for each element $S$ of
$\mathcal{S}_{n-k}$ the relation $ S^2=I_{_{2^n}}$ holds and any
two elements $A_{_{i}}$ and $A_{_{j}}$ of this group satisfy
\begin{equation}\label{staborthogonal}
Tr(A_{_{i}}A_{_{j}})=I_{_{2^n}}\delta_{_{ij}}
\end{equation}
The $n$-qubit Clifford group $Cl(n)$ is the normalizer of
$\mathcal{G}_{n}$ in $U(2^n)$, i.e., it is the group of unitary
operators $U$ satisfying $U\mathcal{G}_{n}U^\dag =
\mathcal{G}_{n}$. It is a finite subgroup of $U(2^n)$ generated by
the Hadamard transform $H$, the phase-shift gate $M$, (both
applied to any single qubit) and the controlled-not gate $CNOT$
which may be applied to any pair of qubits,
$$
\begin{array}{cc}
  H=\frac{1}{\sqrt{2}}\left(\begin{array}{cc}
  1 & 1 \\
  1 & -1 \\
\end{array}
\right),  &\quad M= \left(
\begin{array}{cc}
  1 & 0 \\
  0 & i \\
\end{array}
\right),  \\
\end{array}
$$
$$
CN_{rs}|j\rangle_{r}|k\rangle_{s} = |j\rangle_{r}|j+k\;
\mathrm{mod}\ 2\rangle_{s}\ .
$$
Generators of the Clifford group induce the following
transformations on the Pauli matrices:
\begin{equation}\label{}
    \begin{array}{c}
  H
:\hspace{1cm}\sigma_{_{x}}\longrightarrow\sigma_{_{z}}\quad,
\quad\sigma_{_{y}}\longrightarrow-\sigma_{_{y}}\quad,
\quad\sigma_{_{z}}\longrightarrow\sigma_{_{x}} \\
 M :\hspace{1cm}\sigma_{_{x}}\longrightarrow\sigma_{_{y}}\quad,
\quad\sigma_{_{y}}\longrightarrow-\sigma_{_{x}}\quad,
\quad\sigma_{_{z}}\longrightarrow\sigma_{_{z}} \\
\end{array}
\end{equation}
\begin{equation}\label{}
    CN_{12} : \left\{\begin{array}{ccc}
  I\otimes \sigma_{_{x}} \longrightarrow I\otimes\sigma_{_{x}}, &
  \sigma_{_{x}}\otimes I\longrightarrow\sigma_{_{x}}\otimes\sigma_{_{x}}, &
  \sigma_{_{y}}\otimes \sigma_{_{y}}\longrightarrow -\sigma_{_{x}}\otimes \sigma_{_{z}} \\
  I\otimes\sigma_{_{y}}\longrightarrow -\;\sigma_{_{z}}\otimes\sigma_{_{y}}, &
  \sigma_{_{y}}\otimes I\longrightarrow\sigma_{_{y}}\otimes\sigma_{_{x}}, &
  \sigma_{_{x}}\otimes\sigma_{_{y}}\longrightarrow\sigma_{_{y}}\otimes \sigma_{_{z}} \\
  I\otimes \sigma_{_{z}} \longrightarrow \sigma_{_{z}}\otimes\sigma_{_{z}}, &
  \sigma_{_{z}}\otimes I\longrightarrow\sigma_{_{z}}\otimes I,  &
  \sigma_{_{z}}\otimes\sigma_{_{x}}\longrightarrow\sigma_{_{z}}\otimes \sigma_{_{x}} \\
\end{array}\right.
\end{equation}
and their actions on the eigenvectors of Pauli operators are
\begin{equation}\label{cliffproperty}
    \begin{array}{c}
  H |x^\pm\rangle=|z^\pm\rangle \\
  M |x^\pm\rangle=|y^\pm\rangle\\
  MH |z^\pm\rangle=|y^\pm\rangle,\\
 \end{array}
\end{equation}
In the following table, we give some examples of stabilizer groups
together with the corresponding stabilized states (the states
which are invariant under the action of the stabilizer group):
{\small
\begin{tabular}{|c|c|}
  \hline
  Stabilized state & Generators of stabilizer group \\
  \hline
  $|\psi_{_{00...0}}\rangle$ & $\begin{array}{c}
                               S_{_{1}}^{(\mathrm{GHZ})}:=\sigma_{x}^{(1)}\sigma_{x}^{(2)}...\sigma_{x}^{(n)} \\
                                  S_{_{k}}^{(\mathrm{GHZ})}:=\sigma_{z}^{(k-1)}\sigma_{z}^{(k)}\hspace{1cm}
                                  k=2,3,...,n \\
                               \end{array}$\\ \hline
  $|\mathrm{C}_{n}\rangle=\frac{1}{\sqrt{2^n}}\bigotimes_{a=1}^n
 \left(|0\rangle_{a}+|1\rangle_{a}\;\sigma_{z}^{(a+1)}\right)$ &  $\begin{array}{c}
                                                                   S_{_{1}}^{(\mathrm{C})}=\sigma_{x}^{(1)}\sigma_{z}^{(2)} \\
                                                                     S_{_{k}}^{(\mathrm{C})}=\sigma_{z}^{(k-1)}\sigma_{x}^{(k)}\sigma_{z}^{(k+1)}\hspace{1cm}
                                                                     k=2,3,...,n-1 \\
                                                                     S_{_{n}}^{(\mathrm{C})}=\sigma_{z}^{(n-1)}\sigma_{x}^{(n)} \\
                                                                   \end{array}$\\ \hline
  $\begin{array}{c}
    |\Psi_{1}^{(\mathrm{Fi})}\rangle=\frac{1}{4}\sum_{_{S\in
  \mathcal{S}_{_{Fi}}}}S|00000\rangle \\
    |\Psi_{2}^{(\mathrm{Fi})}\rangle=\frac{1}{4}\sum_{_{S\in
  \mathcal{S}_{_{Fi}}}}S|11111\rangle \\
  \end{array}$ & $\begin{array}{c}
     S_{_{1}}^{(\mathrm{Fi})}=\sigma_{x}^{(1)}\sigma_{z}^{(2)}\sigma_{z}^{(3)}\sigma_{x}^{(4)}\\
     S_{_{2}}^{(\mathrm{Fi})}=\sigma_{x}^{(2)}\sigma_{z}^{(3)}\sigma_{z}^{(4)}\sigma_{x}^{(5)}\\
     S_{_{3}}^{(\mathrm{Fi})}=\sigma_{x}^{(1)}\sigma_{x}^{(3)}\sigma_{z}^{(4)}\sigma_{z}^{(5)}\\
     S_{_{4}}^{(\mathrm{Fi})}=\sigma_{z}^{(1)}\sigma_{x}^{(2)}\sigma_{x}^{(4)}\sigma_{z}^{(5)}\\
                  \end{array}$\\ \hline
  $\begin{array}{c}|\Psi_{ev}^{(\mathrm{Se})}\rangle=\frac{1}{\sqrt{8}}\sum_{_{|\psi\rangle\in
  \mathcal{E}}}|\psi\rangle\\
  |\Psi_{od}^{(\mathrm{Se})}\rangle=\frac{1}{\sqrt{8}}\sum_{_{|\psi\rangle\in
   \mathcal{O}}}|\psi\rangle\\
  \end{array}$ & $\begin{array}{c}
S_{_{1}}^{(\mathrm{Se})}=\sigma_{z}^{(1)}\sigma_{z}^{(3)}\sigma_{z}^{(5)}\sigma_{z}^{(7)}\\
S_{_{2}}^{(\mathrm{Se})}=\sigma_{z}^{(2)}\sigma_{z}^{(3)}\sigma_{z}^{(6)}\sigma_{z}^{(7)}\\
S_{_{3}}^{(\mathrm{Se})}=\sigma_{z}^{(4)}\sigma_{z}^{(5)}\sigma_{z}^{(6)}\sigma_{z}^{(7)}\\
S_{_{4}}^{(\mathrm{Se})}=\sigma_{x}^{(1)}\sigma_{x}^{(3)}\sigma_{x}^{(5)}\sigma_{x}^{(7)}\\
S_{_{5}}^{(\mathrm{Se})}=\sigma_{x}^{(2)}\sigma_{x}^{(3)}\sigma_{x}^{(6)}\sigma_{x}^{(7)}\\
S_{_{6}}^{(\mathrm{Se})}=\sigma_{x}^{(4)}\sigma_{x}^{(5)}\sigma_{x}^{(6)}\sigma_{x}^{(7)}\\
                 \end{array}$\\ \hline
  $\begin{array}{c}
    |\Psi_{_{i_{_{1}}i_{_{2}}i_{_{3}}}}^{(\mathrm{Ei})}\rangle
  =\big(X_{_{1}}\big)^{i_{_{1}}}\big(X_{_{2}}\big)^{i_{_{2}}}\big(X_{_{3}}\big)^{i_{_{3}}}\sum_{_{S\in
  \mathcal{S}_{_{Ei}}}}S|0\rangle^{\otimes 8} \\
    (i_{_{1}},i_{_{2}},i_{_{3}})\in\{0,1\}^{3} \\
  \end{array}$ & $\begin{array}{c}
          S_{_{1}}^{(\mathrm{Ei})}=\sigma_{x}^{(1)}\sigma_{x}^{(2)}...\sigma_{x}^{(8)}\\
          S_{_{2}}^{(\mathrm{Ei})}=\sigma_{z}^{(1)}\sigma_{z}^{(2)}...\sigma_{z}^{(8)}\\
          S_{_{3}}^{(\mathrm{Ei})}=\sigma_{z}^{(1)}\sigma_{x}^{(2)}\sigma_{y}^{(3)}\sigma_{z}^{(5)}\sigma_{x}^{(6)}\sigma_{y}^{(7)}\\
          S_{_{4}}^{(\mathrm{Ei})}=\sigma_{z}^{(2)}\sigma_{z}^{(3)}\sigma_{x}^{(4)}\sigma_{x}^{(5)}\sigma_{y}^{(6)}\sigma_{y}^{(7)}\\
          S_{_{5}}^{(\mathrm{Ei})}=\sigma_{x}^{(1)}\sigma_{x}^{(2)}\sigma_{z}^{(4)}\sigma_{y}^{(5)}\sigma_{y}^{(6)}\sigma_{z}^{(7)}\\
      \end{array}$\\ \hline
  $|\Psi_{{\pm}}^{(\mathrm{Ni})}\rangle:=\left(\frac{1}{\sqrt{2}}(|000\rangle
\pm|111\rangle)\right)^{\otimes 3}$ & $\begin{array}{c}
S_{_{1}}^{(\mathrm{Ni})}=\sigma_{x}^{(1)}\sigma_{x}^{(2)}...\sigma_{x}^{(6)}\quad,\quad
S_{_{2}}^{(\mathrm{Ni})}=\sigma_{x}^{(4)}\sigma_{x}^{(5)}...\sigma_{x}^{(9)}\\
S_{_{3}}^{(\mathrm{Ni})}=\sigma_{z}^{(1)}\sigma_{z}^{(2)}\quad,\quad
S_{_{4}}^{(\mathrm{Ni})}=\sigma_{z}^{(2)}\sigma_{z}^{(3)}\\
S_{_{5}}^{(\mathrm{Ni})}=\sigma_{z}^{(4)}\sigma_{z}^{(5)}\quad,\quad
S_{_{6}}^{(\mathrm{Ni})}=\sigma_{z}^{(5)}\sigma_{z}^{(6)}\\
S_{_{7}}^{(\mathrm{Ni})}=\sigma_{z}^{(7)}\sigma_{z}^{(8)}\quad,\quad
S_{_{8}}^{(\mathrm{Ni})}=\sigma_{z}^{(8)}\sigma_{z}^{(9)}\\
                                       \end{array}$\\  \hline

$|\varphi\rangle=\frac{1}{\sqrt{2}}\left(|v\rangle+S_{_{1}}|v\rangle\right)$&
$S_{_{1}}^{(\mathrm{\varphi})}=S_{_{1}}^{(\mathrm{GHZ})}\quad,\quad S_{_{2}}^{(\mathrm{\varphi})}=\bigotimes_{j=1}^{m}S_{_{2j}}^{(\mathrm{GHZ})}$\\
\hline
\end{tabular}}

where
$$
\begin{array}{c}
\mathcal{E}:=\{|0000000\rangle,|1010101\rangle,|0110011\rangle,|1101001\rangle,|0001111\rangle,|1100110\rangle,|1011010\rangle,|0111100\rangle\},\\
\mathcal{O}:=\{|1111111\rangle,|1110000\rangle,|0100101\rangle,|1000011\rangle,|0010110\rangle,|0101010\rangle,|1001100\rangle,|0011001\rangle\},\\
\hspace{-3.6cm}X_{_{1}}=\sigma_{x}^{(1)}\sigma_{x}^{(2)}\sigma_{z}^{(6)}\sigma_{z}^{(8)}\quad,
\quad
X_{_{2}}=\sigma_{x}^{(1)}\sigma_{x}^{(3)}\sigma_{z}^{(4)}\sigma_{z}^{(7)}\quad,
\quad X_{_{3}}=\sigma_{x}^{(1)}\sigma_{z}^{(4)}\sigma_{x}^{(5)}\sigma_{z}^{(6)},\\
 |v\rangle=
\left(\sigma_{x}^{(1)}\right)^{i_{1}}\left(\sigma_{x}^{(2)}\right)^{i_{2}}...\left(\sigma_{x}^{(2m)}\right)^{i_{2m}}
|0\rangle_{1}|0\rangle_{2}...|0\rangle_{2m},\quad
\oplus_{k=1}^{2m} i_{k}=0, \quad \forall \
(i_{1},i_{2},...,i_{2m})\in \{0,1\}^{2m}
\end{array}
$$
Here $\oplus$ is the sum module 2.
\par
Each of the stabilizer groups stated in the above table,
corresponds to some graph states \cite{guhne4,moor4}. These states
are defined as follows: A graph is a set of $n$ vertices and some
edges connecting them. For every graph $G$, it is associated an
adjacency matrix $T$ whose entries are $T_{ij}=1$ if the vertices
$i$ and $j$ are connected and $T_{ij}=0$ otherwise. Based on that
one can attach a stabilizer operator for every vertex $i$ as
follows
$$
S_{_{i}}^{(G_{n})}=\sigma_{x}^{(i)}\prod_{j\neq i}
{\big(\sigma_{z}^{(j)}\big)}^{T_{ij}}
$$
The graph state $|G\rangle$ associated with the graph $G$ is the
unique n-qubit state satisfying
$$
S_{_{i}}^{(G_{n})}|G\rangle=|G\rangle,\qquad \mathrm{for}\;\;
i=1,...,n.
$$
The case $k=0$ and $k>0$ are called graph state  and graph code
respectively \cite{gott2,grass1}. One can denote the generators of
any stabilizer group by a binary $(n-k)\times 2n$ stabilizer
matrix $[\mathcal{X}|\mathcal{Z}]$ where $\mathcal{X}$ and
$\mathcal{Z}$ are both  $(n-k)\times n$ matrices. Matrices
$\mathcal{X}$ and $\mathcal{Z}$ have a 1 whenever the generator
has a $\sigma_{x}$ and  $\sigma_{z}$ in the appropriate place
respectively. For instance, in the five-qubit case, this form
becomes
$$
[\mathcal{X}|\mathcal{Z}]=\left(\begin{array}{ccccc|ccccc}
  1 & 0 & 0 & 1 & 0 & 0 & 1 & 1 & 0 & 0 \\
  0 & 1 & 0 & 0 & 1 & 0 & 0 & 1 & 1 & 0 \\
  1 & 0 & 1 & 0 & 0 & 0 & 0 & 0 & 1 & 1 \\
  0 & 1 & 0 & 1 & 0 & 1 & 0 & 0 & 0 & 1 \\
\end{array}\right)
$$
By adding $k$ rows to the stabilizer matrix
$[\mathcal{X}|\mathcal{Z}]$ such that the $n$ resulting rows are
linearly independent, one can construct the matrix $\Gamma$
(called generating matrix in coding theory) as follows
$$
\Gamma=\left(\begin{array}{c|c}
         \mathcal{X} & \mathcal{Z} \\ \hline
         \tilde{\mathcal{X}} & \tilde{\mathcal{Z}} \\
       \end{array}\right)
$$
For five-qubit in hand, we have
$$
\Gamma_{5}=\left(\begin{array}{ccccc|ccccc}
  1 & 0 & 0 & 1 & 0 & 0 & 1 & 1 & 0 & 0 \\
  0 & 1 & 0 & 0 & 1 & 0 & 0 & 1 & 1 & 0 \\
  1 & 0 & 1 & 0 & 0 & 0 & 0 & 0 & 1 & 1 \\
  0 & 1 & 0 & 1 & 0 & 1 & 0 & 0 & 0 & 1 \\ \hline
  0 & 0 & 0 & 0 & 0 & 1 & 1 & 1 & 1 & 1 \\
\end{array}\right)
$$
It is necessary to note that the added $k$ rows are not unique and
this freedom in choice leads to the several locally unitary
equivalent graphs for a given graph code.
\par
 By using the Gaussian elimination method on matrix $\Gamma$
one can transform it to the standard form $\Gamma'=[I|A]$, where
$I$ is a $n\times n$ identity matrix and $A\equiv T$ is adjacency
matrix for the related graph G. The standard form of $\Gamma_{5}$
becomes
$$
\Gamma_{5}=\left(\begin{array}{ccccc|ccccc}
  1 & 0 & 0 & 0 & 0 & 0 & 0 & 0 & 1 & 1 \\
  0 & 1 & 0 & 0 & 0 & 0 & 0 & 1 & 1 & 1 \\
  0 & 0 & 1 & 0 & 0 & 0 & 1 & 0 & 0 & 1 \\
  0 & 0 & 0 & 1 & 0 & 1 & 1 & 0 & 0 & 1 \\
  0 & 0 & 0 & 0 & 1 & 1 & 1 & 1 & 1 & 0 \\
\end{array}\right)
$$
The  graphs for the stabilizer groups stated in the above table
are shown in Fig. 2.

\setcounter{equation}{0}
\renewcommand{\theequation}{II-\arabic{equation}}

{\Large{Appendix II:}}\\
{\bf Proving the inequalities}: \\
For simplicity in the following proofs  we introduce the
abbreviations
$$
\alpha_{i}:=\langle\sigma_{x}^{(i)}\rangle\quad\quad\beta_{i}:=\langle\sigma_{y}^{(i)}\rangle\quad\quad
\gamma_{i}:=\langle\sigma_{z}^{(i)}\rangle
$$
\begin{equation}\label{exval}
  \alpha_{i}^2+\beta_{i}^2+\gamma_{i}^2=1
\end{equation}
where $\langle\sigma_{j}^{(i)}\rangle$ with $j=x,y,z$ are the
expectation values of Pauli operators on any arbitrary pure qubit
state.\\
{\bf The proof of (\ref{ineghzev})}:\\
 We give the proof only for the $j=2$ since the proof
for other cases is similar.
$$
|P_{_{1}}\pm P_{_{2}}+\sum_{k=1}^{n'}(-1)^{i_{k}}P_{_{1,2k}}|\leq
|P_{_{1}}|+| P_{_{2}}|+\sum_{k=1}^{n'}|P_{_{1,2k}}|\leq
$$
$$
|\alpha_{_{1}}\alpha_{_{2}}|\left(|\alpha_{_{3}}...\alpha_{_{n}}|+
\sum_{k=2}^{n'}|\alpha_{_{3}}\alpha_{_{4}}...\alpha_{_{2k-2}}
\beta_{_{1,2k-1}}\beta_{_{1,2k}}\alpha_{_{2k+1}}...\alpha_{_{n}}|\right)
+|\beta_{_{1}}\beta_{_{2}}\alpha_{_{3}}...\alpha_{_{n}}|+|\gamma_{_{1}}\gamma_{_{2}}|\leq
$$
$$
|\alpha_{_{1}}\alpha_{_{2}}|
+|\beta_{_{1}}\beta_{_{2}}\alpha_{_{3}}...\alpha_{_{n}}|+|\gamma_{_{1}}\gamma_{_{2}}|\leq
|\alpha_{_{1}}\alpha_{_{2}}|
+|\beta_{_{1}}\beta_{_{2}}|+|\gamma_{_{1}}\gamma_{_{2}}|\leq1
$$
The last inequality follows from the Cauchy-Schwartz inequality
and (\ref{exval}).
\par
To show that the big bracket is smaller than one,
 we write
$$
\left(|\alpha_{_{3}}...\alpha_{_{n}}|+
\sum_{k=2}^{n'}|\alpha_{_{3}}\alpha_{_{4}}...\alpha_{_{2k-2}}
\beta_{_{1,2k-1}}\beta_{_{1,2k}}\alpha_{_{2k+1}}...\alpha_{_{n}}|\right)
$$
$$
=|\alpha_{_{3}}...\alpha_{_{n}}|+|\beta_{_{3}}\beta_{_{4}}\alpha_{_{5}}...\alpha_{_{n}}|
+|\alpha_{_{3}}\alpha_{_{4}}\beta_{_{5}}\beta_{_{6}}\alpha_{_{7}}...\alpha_{_{n}}|
+|\alpha_{_{3}}\alpha_{_{4}}\alpha_{_{5}}\alpha_{_{6}}\beta_{_{7}}\beta_{_{8}}\alpha_{_{9}}...\alpha_{_{n}}|
$$
$$
+...+|\alpha_{_{3}}\alpha_{_{4}}\alpha_{_{5}}...\alpha_{_{2n'-2}}\beta_{_{2n'-1}}\beta_{_{2n'}}|\leq
|\beta_{_{3}}\beta_{_{4}}\alpha_{_{5}}...\alpha_{_{n}}|+|\alpha_{_{3}}\alpha_{_{4}}|
$$
$$
\times\left[|\alpha_{_{5}}...\alpha_{_{n}}|
+|\beta_{_{5}}\beta_{_{6}}\alpha_{_{7}}...\alpha_{_{n}}|
+|\alpha_{_{5}}\alpha_{_{6}}\beta_{_{7}}\beta_{_{8}}\alpha_{_{9}}...\alpha_{_{n}}|+...
+|\alpha_{_{5}}\alpha_{_{6}}...\alpha_{_{2n'-2}}\beta_{_{2n'-1}}\beta_{_{2n'}}|\right]
$$
$$
\leq|\beta_{_{3}}\beta_{_{4}}\alpha_{_{5}}...\alpha_{_{n}}|+|\alpha_{_{3}}\alpha_{_{4}}|
[|\beta_{_{5}}\beta_{_{6}}\alpha_{_{7}}...\alpha_{_{n}}|
$$
$$
+|\alpha_{_{5}}\alpha_{_{6}}|
[|\alpha_{_{7}}\alpha_{_{8}}|[...[|\beta_{_{2n'-3}}\beta_{_{2n'-2}}\alpha_{_{2n'-1}}
\alpha_{_{2n'}}|+|\alpha_{_{2n'-3}}
\alpha_{_{2n'-2}}|[|\alpha_{_{2n'-1}}
\alpha_{_{2n'}}|+|\beta_{_{2n'-1}}\beta_{_{2n'}}]]...]]]
$$
By the Cauchy-Schwartz inequality, the last bracket
$[|\alpha_{_{2n'-1}}
\alpha_{_{2n'}}|+|\beta_{_{2n'-1}}\beta_{_{2n'}}]$ is less than or
equal to 1. Replacing it with its maximum value 1, the same
argument holds for the term
$[|\beta_{_{2n'-3}}\beta_{_{2n'-2}}\alpha_{_{2n'-1}}
\alpha_{_{2n'}}|+|\alpha_{_{2n'-3}} \alpha_{_{2n'-2}}|]$.
Proceeding in this way, we deduce that the big bracket is less
than or equal to 1.
\\
{\bf The proof of (\ref{ineclev})}:
$$
|\gamma_{_{2m-2}}\alpha_{_{2m-1}}\gamma_{_{2m}}\pm\gamma_{_{2m-3}}\alpha_{_{2m-2}}\gamma_{_{2m-1}}+
\gamma_{_{2m-2}}\beta_{_{2m-1}}\beta_{_{2m}} \gamma_{_{2m+1}}|\leq
$$
$$
|\gamma_{_{2m-2}}\alpha_{_{2m-1}}\gamma_{_{2m}}|+|\gamma_{_{2m-3}}\alpha_{_{2m-2}}\gamma_{_{2m-1}}|+
|\gamma_{_{2m-2}}\beta_{_{2m-1}}\beta_{_{2m}} \gamma_{_{2m+1}}|
$$
taking $\gamma_{_{2m-3}}=\gamma_{_{2m+1}}=1$
$$
\leq|\gamma_{_{2m-2}}\alpha_{_{2m-1}}\gamma_{_{2m}}|+|\alpha_{_{2m-2}}\gamma_{_{2m-1}}|+
|\gamma_{_{2m-2}}\beta_{_{2m-1}}\beta_{_{2m}}|
$$
Cauchy-Schwartz inequality yields
$$
\leq\left(\alpha_{_{2m-1}}^2+\gamma_{_{2m-1}}^2+\beta_{_{2m-1}}^2\right)^{\frac{1}{2}}\left(\gamma_{_{2m-2}}^2\gamma_{_{2m}}^2+\alpha_{_{2m-2}}^2
+\gamma_{_{2m-2}}^2\beta_{_{2m}}^2\right)^{\frac{1}{2}}
$$
$$
\leq\left(\gamma_{_{2m-2}}^2(\gamma_{_{2m}}^2+\beta_{_{2m}}^2)+\alpha_{_{2m-2}}^2\right)^{\frac{1}{2}}\leq1
$$
In the above inequalities we have used (\ref{exval}) and the fact
that
\begin{equation}\label{exval1}
  a\alpha_{i}^2+b\beta_{i}^2+c\gamma_{i}^2\leq1
\end{equation}
when the positive coefficients $a,b,c$ are less than or equal to
one. The
proof for even case is the same as the odd case.\\
{\bf The proof of (\ref{ineapprev})}:\\
We prove only the case $l=1$.
$$
|P_{_{j}}+P_{_{1}}+ \sum_{k=1}^{n'}P_{_{1,2k}}+P_{_{1,3}}|
$$
$$
 =|\cos(\theta_{j-1})\cos(\theta_{j})+
\sin(\theta_{1})\sin(\theta_{2})...\sin(\theta_{n})
\{\cos(\phi_{1})\cos(\phi_{2})...\cos(\phi_{n})
$$
$$
+\sum_{k=1}^{n'}\cos(\phi_{1})\cos(\phi_{2})...\cos(\phi_{2k-2})
\sin(\phi_{2k-1})\sin(\phi_{2k})\cos(\phi_{2k+1})...\cos(\phi_{n})
$$
$$
+\cos(\phi_{1})\sin(\phi_{2})\sin(\phi_{3})\cos(\phi_{4})...\cos(\phi_{n})\}|
$$
$$
\leq|\cos(\theta_{j-1})\cos(\theta_{j})|+|\sin(\theta_{j-1})\sin(\theta_{j})|
\times|\cos(\phi_{1})\cos(\phi_{2})...\cos(\phi_{n})
$$
$$
+\sum_{k=1}^{n'}\cos(\phi_{1})\cos(\phi_{2})...\cos(\phi_{2k-2})
\sin(\phi_{2k-1})\sin(\phi_{2k})\cos(\phi_{2k+1})...\cos(\phi_{n})
$$
$$
+\cos(\phi_{1})\sin(\phi_{2})\sin(\phi_{3})\cos(\phi_{4})...\cos(\phi_{n})|
$$
On the other hand, we note that
$$
\cos(\phi_{1})\cos(\phi_{2})...\cos(\phi_{n})
+\sum_{k=1}^{n'}\cos(\phi_{1})\cos(\phi_{2})...\cos(\phi_{2k-2})
\sin(\phi_{2k-1})\sin(\phi_{2k})\cos(\phi_{2k+1})...\cos(\phi_{n})
$$
$$
+\cos(\phi_{1})\sin(\phi_{2})\sin(\phi_{3})\cos(\phi_{4})...\cos(\phi_{n})\leq\frac{1+\sqrt{2}}{2}
$$
Hence we get
$$
|P_{_{j}}+P_{_{1}}+ \sum_{k=1}^{n'}P_{_{1,2k}}+P_{_{1,3}}|\leq
|\cos(\theta_{j-1})\cos(\theta_{j})|+\frac{1+\sqrt{2}}{2}|\sin(\theta_{j-1})\sin(\theta_{j})|\leq
\frac{1+\sqrt{2}}{2}
$$
{\bf The proof of (\ref{ineclusterapp})}:\\
The proof is for $m\geq2$ and the half-spaces with positive
coefficients. The proofs for the other cases are similar.
$$
|P_{2m}+P_{2m-1}+P_{2m,2m+1}+P_{2m-1,2m}|=
$$
$$
|\gamma_{_{2m-1}}\alpha_{_{2m}}\gamma_{_{2m+1}}+\gamma_{_{2m-2}}\alpha_{_{2m-1}}\gamma_{_{2m}}+
\gamma_{_{2m-1}}\beta_{_{2m}}\beta_{_{2m+1}}
\gamma_{_{2m+2}}+\gamma_{_{2m-2}}\beta_{_{2m-1}}\beta_{_{2m}}
\gamma_{_{2m+1}}|\leq
$$
$$
|\cos(\theta_{2m-1})\sin(\theta_{2m})\cos(\phi_{2m})\cos(\theta_{2m+1})+
\cos(\theta_{2m-2})\sin(\theta_{2m-1})\cos(\phi_{2m-1})\cos(\theta_{2m})
$$
$$
+\cos(\theta_{2m-2})\sin(\theta_{2m-1})\sin(\phi_{2m-1})\sin(\theta_{2m})\sin(\phi_{2m})\cos(\theta_{2m+1})
$$
$$
+\cos(\theta_{2m-1})\sin(\theta_{2m})\sin(\phi_{2m})\sin(\theta_{2m+1})\sin(\phi_{2m+1})\cos(\theta_{2m+2})|
$$
We note that the maximum value of the right-hand side of the above
statement is $\frac{2}{\sqrt{3}}$\;. Hence we get
$$
|P_{2m}+P_{2m-1}+P_{2m,2m+1}+P_{2m-1,2m}|\leq \frac{2}{\sqrt{3}}
$$

\setcounter{equation}{0}
 \renewcommand{\theequation}{III-\arabic{equation}}

{\Large{Appendix III:}}\\
{\bf Odd case of GHZ SEWs}\\
 Let us consider the Hermitian operator
\begin{equation}\label{witghzod}
    \mathcal{W}_{_{GHZ}}'^{(n)}=a_{_{0}}I_{_{2^n}}+\sum_{k=1}^n
a_{_{k}}S_{_{k}}^{(\mathrm{GHZ})}+\sum_{k=1}^{n''}
a_{_{1,2k+1}}S_{_{1}}^{(\mathrm{GHZ})}S_{_{2k+1}}^{(\mathrm{GHZ})}
\end{equation}
coming from (\ref{witghzev}) by replacing all even terms with odd
ones $S_{_{1}}^{(\mathrm{GHZ})}S_{_{2k+1}}^{(\mathrm{GHZ})}$ (the
name odd  refer to the index $2k+1$). It is easily seen that the
eigenvalues of $\mathcal{W'}_{_{GHZ}}^{(n)}$ are

\begin{equation}\label{eigghzod}
    a_{_{0}}+\sum_{j=1}^n (-1)^{i_{j}}
a_{_{j}}+\sum_{k=1}^{n''}(-1)^{i_{1}+ i_{2k+1}}a_{_{1,2k+1}}\quad
,\quad \forall \ (i_{1},i_{2},...,i_{n})\in \{0,1\}^{n}
\end{equation}
The product vectors and the vertex points of the feasible region
 are listed in the table 8,
\begin{table}[h]
\renewcommand{\arraystretch}{1}
\addtolength{\arraycolsep}{-2pt}
$$
\begin{array}{|c|c|}\hline
\mathrm{Product\ state }&
(P_{2},P_{3},...,P_{n-1},P_{n},P_{1},P_{1,3},P_{1,5},...,P_{1,2n''-1},P_{1,2n''+1})
\\ \hline
  |\Psi^{\pm}\rangle & (0,0,...,0,0,\pm1,0,0,...,0,0) \\
  \Lambda'_{_{1}} |\Psi^{\pm}\rangle & (0,0,...,0,0,0,\pm1,0,...,0,0) \\
  \vdots & \vdots \\
  \Lambda'_{_{n''}} |\Psi^{\pm}\rangle & (0,0,...,0,0,0,0,0,...,0,\pm1) \\
  \hline
  \Xi_{i_{2},...,i_{n}}|\Psi^{+}\rangle & \left((-1)^{i_{_{2}}},
  (-1)^{i_{_{2}}+i_{_{3}}},...,(-1)^{i_{_{n-2}}+i_{_{n-1}}},(-1)^{i_{_{n-1}}+i_{_{n}}},0,0,0,...,0,0\right)\\
     \hline
\end{array}
$$
\caption{The product vectors and coordinates of vertices for
$\mathcal{W'}_{_{GHZ}}^{(n)}$.}
\renewcommand{\arraystretch}{1}
\addtolength{\arraycolsep}{-3pt}
\end{table}
where
$$
\begin{array}{c}
    \Lambda'_{_{k}}=\left({M^{(2k)}}\right)^{\dagger} M^{(2k+1)},\quad\quad k=1,2,...,n''\\
  \end{array}
$$
and $|\Psi^{\pm}\rangle$ is defined as in (\ref{tab1'}). The
convex hull of the above vertices, the feasible region, comes from
the intersection of the half-spaces
\begin{equation}\label{ineghzodd}
 |P_{_{1}}\pm P_{_{j}}+\sum_{k=1}^{n''}(-1)^{i_{k}}P_{_{1,2k+1}}|\leq 1
\quad,\quad j=2,...,n \quad,\quad \forall \
(i_{1},i_{2},...,i_{n''})\in \{0,1\}^{n''}
\end{equation}
and is a $(n-1)2^{n''+2}$-simplex.

We give the proof only for the $j=2$ since the proof for other
cases is similar.
$$
|P_{_{1}}\pm
P_{_{2}}+\sum_{k=1}^{n''}(-1)^{i_{k}}P_{_{1,2k+1}}|\leq
|P_{_{1}}|+| P_{_{2}}|+\sum_{k=1}^{n''}|P_{_{1,2k+1}}|\leq
$$
$$
|\alpha_{_{1}}\alpha_{_{2}}|\left(|\alpha_{_{3}}...\alpha_{_{n}}|+
\sum_{k=2}^{n''}|\alpha_{_{3}}\alpha_{_{4}}...\alpha_{_{2k-1}}
\beta_{_{1,2k}}\beta_{_{1,2k+1}}\alpha_{_{2k+2}}...\alpha_{_{n}}|\right)
+|\beta_{_{1}}\beta_{_{2}}\alpha_{_{3}}...\alpha_{_{n}}|+|\gamma_{_{1}}\gamma_{_{2}}|\leq
$$
$$
|\alpha_{_{1}}\alpha_{_{2}}|
+|\beta_{_{1}}\beta_{_{2}}\alpha_{_{3}}...\alpha_{_{n}}|+|\gamma_{_{1}}\gamma_{_{2}}|\leq
|\alpha_{_{1}}\alpha_{_{2}}|
+|\beta_{_{1}}\beta_{_{2}}|+|\gamma_{_{1}}\gamma_{_{2}}|\leq1
$$
The last inequality follows from the Cauchy-Schwartz inequality
and (\ref{exval}).\\
Now the problem is reduced to the following LP problem
\begin{equation}\label{lpghzod}
\begin{array}{c}
\mathrm{minimize}\quad
\mathcal{F}_{\mathcal{W'}_{_{GHZ}}^{(n)}}=a_{_{0}}+\sum_{k=1}^n
a_{_{k}}P_{{k}}+\sum_{k=1}^{n''}a_{_{1,2k+1}}P_{{1,2k+1}}\\
\mathrm{subject\; to}\quad |P_{_{1}}\pm
P_{_{j}}+\sum_{k=1}^{n''}(-1)^{i_{k}}P_{_{1,2k+1}}|\leq 1 ,\quad
j=2,...,n \quad, \forall \
(i_{1},i_{2},...,i_{n''})\in \{0,1\}^{n''}\\
\end{array}
\end{equation}
By putting  the coordinates of vertices (see table 2) in the
objective function $\mathcal{F}_{\mathcal{W'}_{_{GHZ}}^{(n)}}$ and
requiring its non-negativity on all vertices, we get the
conditions
\begin{equation}\label{ineqparaghzod}
  \begin{array}{c}
   a_{_{0}}>0\quad,\quad a_{_{0}}\geq  |a_{_{1}}|\quad,\quad a_{_{0}}\geq\sum_{i=2}^n |a_{_{i}}|\\
   a_{_{0}}\geq |a_{_{1,2k+1}}| \quad\quad k=1,...,n''\\
  \end{array}
\end{equation}
on parameters $a_i$.\\
{\bf Odd case of cluster SEWs}\\
Let us consider the Hermitian operators
\begin{equation}\label{witclod}
\mathcal{W'}_{_{C}}^{(n)}=a_{_{0}}I_{_{2^n}}+\sum_{k=0}^{n''}
a_{_{2k+1}}S_{_{2k+1}}^{(\mathrm{C})}
+a_{_{2m}}S_{_{2m}}^{(\mathrm{C})}+a_{_{2m,2m+1}}S_{_{2m}}^{(\mathrm{C})}S_{_{2m+1}}^{(\mathrm{C})}\quad,\quad
m=1,...,n''
\end{equation}
Note that instead of the last term we can put the term
$a_{_{2m-1,2m}}S_{_{2m-1}}^{(\mathrm{C})}S_{_{2m}}^{(\mathrm{C})}$
with $m=1,...,n'$.  The eigenvalues of $\mathcal{W'}_{_{C}}^{(n)}$
are
\begin{equation}\label{eigenclod}
a_{_{0}}+\sum_{j=0}^{n''} (-1)^{i_{2j+1}}
a_{_{2j+1}}+(-1)^{i_{2m}}a_{_{2m}}+(-1)^{i_{2m}+i_{2m+1}}a_{_{2m,2m+1}}
\quad, \quad \forall \ (i_{1},i_{2},...,i_{n})\in \{0,1\}^{n}
\end{equation}
 The product vectors and the
vertex points of the feasible region  are listed in the table 9

\begin{table}[h]
\renewcommand{\arraystretch}{1.2}
\addtolength{\arraycolsep}{-2pt}
$$
\begin{array}{|c|c|}\hline
 \mathrm{Product\ state} & (P_{1},P_{3},...,P_{2m-3},P_{2m-1},P_{2m},P_{2m+1},P_{2m+3},..., P_{2n''+1},P_{2m,2m+1}) \\ \hline
  \Lambda_{_{i_{1},i_{2},...,i_{n''+1}}}^{(odd)}|\Phi\rangle & \big((-1)^{i_{1}},(-1)^{i_{2}},...,
  (-1)^{i_{m-2}},(-1)^{i_{m-1}},0,(-1)^{i_{m}},(-1)^{i_{m+1}},...,(-1)^{i_{n''+1}},0\big) \\
   \hline
  {\Lambda'}_{_{i_{1},i_{2},...,i_{n''+1}}}^{(odd)}|\Phi\rangle & \big((-1)^{i_{1}},(-1)^{i_{2}},...,
  (-1)^{i_{m-2}},0,\pm1,0,(-1)^{i_{m+1}},...,(-1)^{i_{n''+1}},0\big)\\
  \hline
  {\Lambda''}_{_{i_{1},i_{2},...,i_{n''+1}}}^{(odd)}|\Phi\rangle & ((-1)^{i_{1}},(-1)^{i_{2}},...,(-1)^{i_{m-2}},0,0,0,(-1)
  ^{i_{m+1}},...,(-1)^{i_{n''+1}},(-1)^{i_{m}})\\
  \hline
\end{array}
$$
\caption{The product vectors and coordinates of vertices for
$\mathcal{W'}_{_{C}}^{(n)}$.}
\renewcommand{\arraystretch}{1}
\addtolength{\arraycolsep}{3pt}
\end{table}
where
$$
\begin{array}{c}
  |\Phi\rangle=|z^+\rangle_{_{1}}|x^+\rangle_{_{2}}|z^+\rangle_{_{3}}|x^+\rangle_{_{4}}|z^+\rangle_{_{5}}...
  |x^+\rangle_{_{n-1}}|z^+\rangle_{_{n}} \\
  \Lambda_{_{i_{1},i_{2},...,i_{n''+1}}}^{(odd)}=\bigotimes_{j=1}^{n''+1}\left(\sigma_{_{z}}^{(2j+1)}\right)^{i_{j}}
 \bigotimes_{j=1}^n H^{(j)}\quad,\quad \forall\ (i_{1},i_{2},...,i_{n''+1})\in \{0,1\}^{n''+1}\\
   {\Lambda'}_{_{i_{1},i_{2},...,i_{n''+1}}}^{(odd)}= \Lambda_{_{i_{1},i_{2},...,i_{n''+1}}}^{(odd)}H^{(2m-1)}H^{(2m)}H^{(2m+1)}\\
  {\Lambda''}_{_{i_{1},i_{2},...,i_{n''+1}}}^{(odd)}= \Lambda_{_{i_{1},i_{2},...,i_{n''+1}}}^{(odd)}
  H^{(2m-1)}M^{(2m)}H^{(2m)}M^{(2m+1)}\\
\end{array}
$$
For a given $m$, the feasible region (the convex hull of the above
vertices),  comes from the intersection of the half-spaces
\begin{equation}\label{ineclod}
\begin{array}{c}
  |P_{2m}\pm P_{2m-1}+P_{2m,2m+1}|\leq1 \\
  |P_{2m}\pm P_{2m-1}-P_{2m,2m+1}|\leq1 \\
  |P_{2m}\pm P_{2m+1}+P_{2m,2m+1}|\leq1 \\
  |P_{2m}\pm P_{2m+1}-P_{2m,2m+1}|\leq1 \\
  \hspace{2cm}|P_{2k+1}|\leq1 \quad,\quad m,m-1\neq k=1,...,n''  \\
\end{array}
\end{equation}
and is a $(2n''+12)$-simplex. For the first two inequalities we
have
$$
|\gamma_{_{2m-1}}\alpha_{_{2m}}\gamma_{_{2m+1}}\pm\gamma_{_{2m-2}}\alpha_{_{2m-1}}\gamma_{_{2m}}+
\gamma_{_{2m-1}}\beta_{_{2m}}\beta_{_{2m+1}} \gamma_{_{2m+2}}|\leq
$$
$$
|\gamma_{_{2m-1}}\alpha_{_{2m}}\gamma_{_{2m+1}}|+|\gamma_{_{2m-2}}\alpha_{_{2m-1}}\gamma_{_{2m}}|+
|\gamma_{_{2m-1}}\beta_{_{2m}}\beta_{_{2m+1}} \gamma_{_{2m+2}}|
$$
taking $\gamma_{_{2m-2}}=\gamma_{_{2m+2}}=1$
$$
\leq|\gamma_{_{2m-1}}\alpha_{_{2k}}\gamma_{_{2m+1}}|+|\alpha_{_{2m-1}}\gamma_{_{2m}}|+|\gamma_{_{2m-1}}\beta_{_{2m}}\beta_{_{2m+1}}
|\leq
$$
Cauchy-Schwartz inequality yields
$$
\leq\left(\alpha_{_{2m}}^2+\gamma_{_{2m}}^2+\beta_{_{2m}}^2\right)^\frac{1}{2}
\left(\gamma_{_{2m-1}}^2\gamma_{_{2m+1}}^2+\alpha_{_{2m-1}}^2
+\gamma_{_{2m-1}}^2\beta_{_{2m+1}}^2\right)^\frac{1}{2}
$$
$$
\leq\left(\gamma_{_{2m-1}}^2(\gamma_{_{2m+1}}^2+\beta_{_{2m+1}}^2)+\alpha_{_{2m-1}}^2\right)^\frac{1}{2}\leq1
$$
In the above inequalities we have used (\ref{exval}) and
(\ref{exval1}).
\par
For the second two inequalities we have
$$
|\gamma_{_{2m-1}}\alpha_{_{2m}}\gamma_{_{2m+1}}\pm\gamma_{_{2m}}\alpha_{_{2m+1}}\gamma_{_{2m+2}}+
\gamma_{_{2m-1}}\beta_{_{2m}}\beta_{_{2m+1}} \gamma_{_{2m+2}}|\leq
$$
$$
|\gamma_{_{2m-1}}\alpha_{_{2m}}\gamma_{_{2m+1}}|+|\gamma_{_{2m}}\alpha_{_{2m+1}}\gamma_{_{2m+2}}|+
|\gamma_{_{2m-1}}\beta_{_{2m}}\beta_{_{2m+1}} \gamma_{_{2m+2}}|
$$
taking $\gamma_{_{2m-1}}=\gamma_{_{2m+2}}=1$
$$
\leq|\alpha_{_{2m}}\gamma_{_{2m+1}}|+|\gamma_{_{2m}}\alpha_{_{2m+1}}|+
|\beta_{_{2m}}\beta_{_{2m+1}} |\leq1
$$
The last inequality follows from the Cauchy-Schwartz inequality.
\par
The objective function is
\begin{equation}\label{objclod}
\mathcal{F}_{\mathcal{W'}_{_{C}}^{(n)}}=a_{_{0}}+\sum_{k=0}^{n''}
a_{_{2k+1}}P_{_{2k+1}}
+a_{_{2m}}P_{_{2m}}+a_{_{2m,2m+1}}P_{_{2m,2m+1}}\quad,\quad
m=1,...,n''
\end{equation}
where
$$
P_{_{2k+1}}=Tr(S_{_{2k+1}}^{(\mathrm{C})}\rho_{s})\quad,\quad
P_{_{2m,2m+1}}=Tr(S_{_{2m}}^{(\mathrm{C})}S_{_{2m+1}}^{(\mathrm{C})}\rho_{s}),
$$

 If we put the coordinates of vertices (see table
4) in the objective function (\ref{objclod}) and require the
non-negativity of the objective function on all vertices, we get
the conditions
\begin{equation}\label{ineqparaclod}
\begin{array}{c}
a_{_{0}} \geq \sum_{j=0}^{n''}|a_{_{2j+1}}| \\
a_{_{0}}\geq\sum_{j=1}^{m-2}|a_{_{2j+1}}|+\sum_{j=m+1}^{n''}|a_{_{2j+1}}|+|a_{_{2m}}|\\
a_{_{0}}\geq\sum_{j=1}^{m-2}|a_{_{2j+1}}|+\sum_{j=m+1}^{n''}|a_{_{2j+1}}|+|a_{_{2m,2m+1}}|\\
\end{array}
\end{equation}
for parameters  $a_i$.\\
{\bf Exceptional SEWs}\\
 Here we mention briefly the SEWs that can
be constructed by the stabilizer operations of the five-qubit,
seven-qubit, eight-qubit, and nine-qubit states that can be solved
by exact LP method.\\
{\bf Five-qubit SEWs} \\
Consider the following Hermitian operator
$$
\mathcal{W}_{_{Fi}}=a_{_{0}}I_{_{2^5}}+ a_{_{1}}S_{_{1}}^{(Fi)}+
a_{_{2}}S_{_{2}}^{(Fi)}+
a_{_{3}}S_{_{3}}^{(Fi)}+a_{_{3,4}}S_{_{3}}^{(Fi)}S_{_{4}}^{(Fi)}
$$
Eigenvalues of $\mathcal{W}_{_{Fi}}$ are
$$
a_{_{0}}+\sum_{j=1}^{3}(-1)^{i_{j}}a_{_{j}}\pm a_{_{3,4}}
\quad,\quad\forall\;(i_{1},...,i_{3})\in\{0,1\}^{3}
$$
The vertex points of the feasible region are listed in Table 10

\begin{table}[h]
\renewcommand{\arraystretch}{1}
\addtolength{\arraycolsep}{2pt}
$$
\begin{array}{|c|c|}\hline
   \mathrm{Product\ state} & (P_{1},P_{2},P_{3},P_{3,4}) \\ \hline
  |\Psi_{_{\pm}}^{(Fi)}\rangle & (\pm1,0,0,0) \\
  H^{(2)}H^{(4)}(\mathrm{SW})_{_{15}} |\Psi_{_{\pm}}^{(Fi)}\rangle & (0,\pm1,0,0) \\
  H^{(3)}H^{(4)}(\mathrm{SW})_{_{25}} |\Psi_{_{\pm}}^{(Fi)}\rangle & (0,0,\pm1,0) \\
  H^{(1)}H^{(2)}(\mathrm{SW})_{_{35}} |\Psi_{_{\pm}}^{(Fi)}\rangle & (0,0,0,\pm1) \\
  \hline
\end{array}
$$
\caption{The product vectors and coordinates of vertices for
$\mathcal{W}_{_{Fi}}$.}
\renewcommand{\arraystretch}{1}
\addtolength{\arraycolsep}{3pt}
\end{table}

where
$$
\begin{array}{c}
 |\Psi_{_{\pm}}^{(Fi)}\rangle=|x^\pm\rangle_{_{1}}|z^+\rangle_{_{2}}|z^+\rangle_{_{3}}|x^+\rangle_{_{4}}|\;\;\rangle_{_{5}}\\
 (\mathrm{SW})_{_{ij}}=(CN)_{ij}(CN)_{ji}(CN)_{ij}\\
   \end{array}
$$
The operator $(SW)_{ij}$ when acts on any two arbitrary pure
states swaps them, i.e.,
$(SW)_{ij}|\psi\rangle_{i}|\varphi\rangle_{j}=|\varphi\rangle_{i}|\psi\rangle_{j}$.
Inequalities obtained from putting the vertex points are
$$
 a_{_{0}}\geq |a_{_{i}}|\quad\quad i=1,2,3\quad,\quad a_{_{0}}\geq |a_{_{3,4}}|
$$
Boundary half-spaces of feasible region are
\begin{equation}\label{inefive}
\begin{array}{c}
  |P_{_{1}}\pm P_{_{2}}+P_{_{3}}+P_{_{3,4}}|\leq 1\quad,\quad |P_{_{1}}\pm P_{_{2}}+P_{_{3}}-P_{_{3,4}}|\leq 1 \\
  |P_{_{1}}\pm P_{_{2}}-P_{_{3}}+P_{_{3,4}}|\leq 1\quad,\quad |P_{_{1}}\pm P_{_{2}}-P_{_{3}}-P_{_{3,4}}|\leq 1 \\
\end{array}
\end{equation}
We prove only the following inequality since the proof of the
other inequalities is similar to this one.
$$
 |P_{_{1}}+P_{_{2}}+P_{_{3}}+P_{_{3,4}}|=|\alpha_{_{1}}\gamma_{_{2}}\gamma_{_{3}}\alpha_{_{4}}
 +\alpha_{_{2}}\gamma_{_{3}}\gamma_{_{4}}\alpha_{_{5}}+
\alpha_{_{1}}\alpha_{_{3}}\gamma_{_{4}}\gamma_{_{5}}+\beta{_{1}}\alpha_{_{2}}\alpha_{_{3}}\beta{_{4}}\leq
$$
$$
|\alpha_{_{1}}\gamma_{_{2}}\gamma_{_{3}}\alpha_{_{4}}|+|\alpha_{_{2}}\gamma_{_{3}}\gamma_{_{4}}\alpha_{_{5}}|+|
\alpha_{_{1}}\alpha_{_{3}}\gamma_{_{4}}\gamma_{_{5}}|+|\beta{_{1}}\alpha_{_{2}}\alpha_{_{3}}\beta{_{4}}|\leq
$$
$$
|\gamma_{_{3}}|(|\alpha_{_{1}}\gamma_{_{2}}\alpha_{_{4}}|+|\alpha_{_{2}}\gamma_{_{4}}\alpha_{_{5}}|)+|
\alpha_{_{3}}|(|\alpha_{_{1}}\gamma_{_{4}}\gamma_{_{5}}|+|\beta{_{1}}\alpha_{_{2}}\beta{_{4}}|)\leq
$$
$$
|\gamma_{_{3}}|\left(\alpha_{_{2}}^2+\gamma_{_{2}}^2\right)^\frac{1}{2}
\left(\alpha_{_{1}}^2\alpha_{_{4}}^2+\gamma_{_{4}}^2\alpha_{_{5}}^2\right)^\frac{1}{2}
+|\alpha_{_{3}}|
\left(\alpha_{_{1}}^2+\beta_{_{1}}^2\right)^\frac{1}{2}
\left(\gamma_{_{4}}^2\gamma_{_{5}}^2+\alpha_{_{2}}^2\beta_{_{4}}^2\right)^\frac{1}{2}\leq
$$
$$
|\gamma_{_{3}}|\left(\alpha_{_{1}}^2\alpha_{_{4}}^2+\gamma_{_{4}}^2\alpha_{_{5}}^2\right)^\frac{1}{2}
+|\alpha_{_{3}}|
\left(\gamma_{_{4}}^2\gamma_{_{5}}^2+\alpha_{_{2}}^2\beta_{_{4}}^2\right)^\frac{1}{2}\leq
$$
$$
\left(\alpha_{_{3}}^2+\gamma_{_{3}}^2\right)^\frac{1}{2}
\left(\alpha_{_{1}}^2\alpha_{_{4}}^2+\gamma_{_{4}}^2\alpha_{_{5}}^2
+
\gamma_{_{4}}^2\gamma_{_{5}}^2+\alpha_{_{2}}^2\beta_{_{4}}^2\right)^\frac{1}{2}\leq
$$
$$
\left(\alpha_{_{1}}^2\alpha_{_{4}}^2+\gamma_{_{4}}^2(\alpha_{_{5}}^2
+
\gamma_{_{5}}^2)+\alpha_{_{2}}^2\beta_{_{4}}^2\right)^\frac{1}{2}\leq
\left(\alpha_{_{1}}^2\alpha_{_{4}}^2+\gamma_{_{4}}^2+\alpha_{_{2}}^2
\beta_{_{4}}^2\right)^\frac{1}{2}\leq\left(\alpha_{_{4}}^2+\gamma_{_{4}}^2+\beta_{_{4}}^2\right)^\frac{1}{2}\leq1
$$
The above inequalities follow from the Cauchy-Schwartz inequality
and the equations (\ref{exval}) and (\ref{exval1}).
\par
 From $2^4$ eigenvalues of
$\mathcal{W}_{_{Fi}}$, six of them can take negative values. If we
take all $a_{_{1}},a_{_{2}},a_{_{3}},a_{_{3,4}}$ positive and
without loss of generality assume that $a_{_{1}}\geq a_{_{2}}\geq
a_{_{3}}\geq a_{_{3,4}}$, then these eigenvalues are
$$
a_{_{0}}-a_{_{1}}-a_{_{2}}\pm a_{_{3}}+a_{_{3,4}}\quad,\quad
a_{_{0}}-a_{_{1}}-a_{_{2}}\pm a_{_{3}}-a_{_{3,4}}\quad,\quad
a_{_{0}}\pm a_{_{1}}\mp a_{_{2}}-a_{_{3}}-a_{_{3,4}}
$$
{\bf Seven-qubit SEWs}\\
Consider the following Hermitian operator
$$
\mathcal{W}_{_{Se}}=a_{_{0}}I_{_{2^7}}+ \sum_{i=1}^6
a_{_{i}}S_{_{i}}^{(Se)}+ a_{_{1,4}}S_{_{1}}^{(Se)}S_{_{4}}^{(Se)}
$$
In addition to the above operator, we can consider other Hermitian
operators which differ from the above operator only in the last
term, that is the last term of them is one of the following
operators
$$
S_{_{1}}^{(Se)}S_{_{4}}^{(Se)},\;S_{_{2}}^{(Se)}S_{_{5}}^{(Se)},\;
S_{_{3}}^{(Se)}S_{_{6}}^{(Se)},\;S_{_{1}}^{(Se)}S_{_{5}}^{(Se)},\;
S_{_{2}}^{(Se)}S_{_{6}}^{(Se)}
$$
Eigenvalues of $\mathcal{W}_{_{Se}}$ are
$$
a_{_{0}}+\sum_{j=1}^{6}(-1)^{i_{j}}a_{_{j}}+(-1)^{i_{1}+i_{4}}a_{_{1,4}}\quad\quad
\forall\;(i_{1},...,i_{6})\in\{0,1\}^{6}
$$
The vertex points of feasible region are listed in table 11

\begin{table}[h]
\renewcommand{\arraystretch}{1.2}
\addtolength{\arraycolsep}{-2pt}
$$
\begin{array}{|c|c|}\hline
 \mathrm{Product\ state} & (P_{1},P_{2},P_{3},P_{4},P_{5},P_{6},P_{1,4}) \\ \hline
  \Lambda_{_{i_{1},i_{2},i_{3}}}^{(\mathrm{Se})}|\Phi^{(\mathrm{Se})}\rangle & \big((-1)^{i_{1}},(-1)^{i_{2}},
  (-1)^{i_{3}},0,0,0,0\big) \\
   \hline
  {\Lambda'}_{_{i_{1},i_{2},i_{3}}}^{(\mathrm{Se})}|\Phi^{(\mathrm{Se})}\rangle & \big(0,0,0,(-1)^{i_{1}},(-1)^{i_{2}},
  (-1)^{i_{3}},0\big)\\
  \hline
  {\Lambda}_{_{i}}^{(\mathrm{Se})}|\Phi^{(\mathrm{Se})}\rangle & \big(0,0,0,0,0,0,(-1)
  ^{i}\big)\\
  \hline
\end{array}
$$
\caption{The product vectors and coordinates of vertices for
$\mathcal{W}_{_{Se}}$.}
\renewcommand{\arraystretch}{1}
\addtolength{\arraycolsep}{3pt}
\end{table}

where
$$
\begin{array}{c}
  |\Phi^{(Se)}\rangle=|z^+\rangle_{_{1}}|z^+\rangle_{_{2}}...|z^+\rangle_{_{7}}\\
  \Lambda_{_{i_{1},i_{2},i_{3}}}^{(\mathrm{Se})}=\big(\sigma_{x}^{(1)}\big)^{i_{_{1}}}\big(\sigma_{x}^{(2)}\big)^{i_{_{2}}}
  \big(\sigma_{x}^{(4)}\big)^{i_{_{3}}}\\
   {\Lambda'}_{_{i_{1},i_{2},i_{3}}}^{(\mathrm{Se})}=\big(\sigma_{z}^{(1)}\big)^{i_{_{1}}}\big(\sigma_{z}^{(2)}\big)^{i_{_{2}}}
  \big(\sigma_{z}^{(4)}\big)^{i_{_{3}}}\bigotimes_{j=1}^7 H^{(j)},\quad\ \forall\ (i_{1},i_{2},i_{3})\in \{0,1\}^{3} \\
  {\Lambda}_{_{i}}^{(\mathrm{Se})}=\big(\sigma_{z}^{(1)}\big)^{i}\bigotimes_{j=1}^4 M^{(2j-1)}H^{(2j-1)}
  ,\quad\ \forall\ i\in \{0,1\}\\
\end{array}
$$
Boundary half-spaces of feasible region are
\begin{equation}\label{ineseven}
 \begin{array}{c}
 |P_{_{i}}\pm P_{_{j}}+P_{_{1,4}}|\leq 1\quad,\quad |P_{_{i}}\pm P_{_{j}}-P_{_{1,4}}|\leq 1
 \quad\quad i=1,2,3 \quad,\quad
 j=4,5,6 \\
\end{array}
\end{equation}
Although all of the inequalities (\ref{ineseven}) can be derived
by Cauchy-Schwartz inequality but require a tricky way. The proof
of two cases $i=2,j=6$ and $i=3,j=5$  are similar and therefore we
prove only the former case.
$$
 |P_{_{2}}+P_{_{6}}+P_{_{1,4}}|=|\gamma_{_{2}}\gamma_{_{3}}\gamma_{_{6}}\gamma_{_{7}}
+\alpha_{_{4}}\alpha_{_{5}}\alpha_{_{6}}\alpha_{_{7}}+\beta_{_{1}}\beta_{_{3}}\beta_{_{5}}\beta_{_{7}}|\leq
$$
$$
 |\gamma_{_{2}}\gamma_{_{3}}\gamma_{_{6}}\gamma_{_{7}}|
+|\alpha_{_{4}}\alpha_{_{5}}\alpha_{_{6}}\alpha_{_{7}}|+|\beta_{_{1}}\beta_{_{3}}\beta_{_{5}}\beta_{_{7}}|
$$
taking $\gamma_{_{2}}=\alpha_{_{4}}=\beta_{_{1}}=1$
$$
 \leq|\gamma_{_{3}}\gamma_{_{6}}\gamma_{_{7}}|
+|\alpha_{_{5}}\alpha_{_{6}}\alpha_{_{7}}|+|\beta_{_{3}}\beta_{_{5}}\beta_{_{7}}|\leq
$$
$$
\big(\alpha_{_{7}}^2+\beta_{_{7}}^2+\gamma_{_{7}}^2\big)^{\frac{1}{2}}
\big(\gamma_{_{3}}^2\gamma_{_{6}}^2
+\alpha_{_{5}}^2\alpha_{_{6}}^2+\beta_{_{3}}^2\beta_{_{5}}^2\big)^{\frac{1}{2}}=
$$
$$
\big[\gamma_{_{3}}^2\gamma_{_{6}}^2(\alpha_{5}^2+\beta_{5}^2+\gamma_{5}^2)
+\alpha_{_{5}}^2\alpha_{_{6}}^2(\alpha_{3}^2+\beta_{3}^2+\gamma_{3}^2)
+\beta_{_{3}}^2\beta_{_{5}}^2(\alpha_{6}^2+\beta_{6}^2+\gamma_{6}^2)\big]^{\frac{1}{2}}\leq
$$
$$
\big[(\alpha_{3}^2+\beta_{3}^2+\gamma_{3}^2)(\alpha_{5}^2+\beta_{5}^2+
\gamma_{5}^2)(\alpha_{6}^2+\beta_{6}^2+\gamma_{6}^2)\big]^{\frac{1}{2}}=1
$$
The above inequalities follow from the Cauchy-Schwartz inequality
and the equations (\ref{exval}) and (\ref{exval1}).
\par
Inequalities obtained from putting of vertex points are
$$
 \begin{array}{c}
  a_{_{0}} \geq |a_{_{1,4}}|\quad,\quad a_{_{0}}\geq\sum_{j=1}^{3}|a_{_{j}}|
  \quad,\quad a_{_{0}}\geq\sum_{j=4}^{6}|a_{_{j}}|  \\
\end{array}
$$
{\bf Eight-qubit SEWs}\\
 Consider the following Hermitian operator
$$
\mathcal{W}_{_{\mathrm{Ei}}}=a_{_{0}}I_{_{2^8}}+\sum_{i=1}^{5}
a_{_{i}}S_{_{i}}^{(\mathrm{Ei})}+
a_{_{1,2,3}}S_{_{1}}^{(\mathrm{Ei})}S_{_{2}}^{(\mathrm{Ei})}S_{_{3}}^{(\mathrm{Ei})}+
a_{_{1,2,4}}S_{_{1}}^{(\mathrm{Ei})}S_{_{2}}^{(\mathrm{Ei})}S_{_{4}}^{(\mathrm{Ei})}
$$
Eigenvalues of $\mathcal{W}_{_{Ei}}$ are
$$
a_{_{0}}+ \sum_{j=1}^{5}(-1)^{i_{j}}
a_{_{j}}+(-1)^{i_{1}+i_{2}+i_{3}}a_{_{1,2,3}}+(-1)^{i_{1}+i_{2}+i_{4}}a_{_{1,2,4}}
\quad\quad\forall\;(i_{1},i_{2},...,i_{5})\in\{0,1\}^{5}
$$
The vertex points of feasible region are listed in table 12

\begin{table}[h]
\renewcommand{\arraystretch}{1.2}
\addtolength{\arraycolsep}{0pt}
$$
\begin{array}{|c|c|}\hline
 \mathrm{Product\ state} & (P_{_{1}},P_{_{2}},P_{_{3}},P_{_{4}}, P_{_{5}},P_{_{1,2,3}}, P_{_{1,2,4}})\\
  \hline
  |\Phi_{_{\pm}}^{(\mathrm{Ei})}\rangle & \big(\pm1,0,
  0,0,0,0,0\big) \\
   \hline
  H^{(1)}H^{(2)}...H^{(8)}|\Phi_{_{\pm}}^{(\mathrm{Ei})}\rangle & \big(0,\pm1,
  0,0,0,0,0\big) \\
  \hline
  H^{(1)}M^{(3)}H^{(5)}M^{(7)}|\Phi_{_{\pm}}^{(\mathrm{Ei})}\rangle & \big(0,0,
  \pm1,0,0,0,0\big)\\
  \hline
  H^{(2)}H^{(3)}M^{(6)}M^{(7)}|\Phi_{_{\pm}}^{(\mathrm{Ei})}\rangle &
  \big(0,0,0,\pm1,0,0,0\big)\\
  \hline
 H^{(4)}M^{(5)}M^{(6)}H^{(7)}|\Phi_{_{\pm}}^{(\mathrm{Ei})}\rangle &
  \big(0,0,0,0,\pm1,0,0\big)\\
  \hline
  H^{(2)}M^{(4)}M^{(8)}|\Phi_{_{\pm}}^{(\mathrm{Ei})}\rangle &
  \big(0,0,0,0,0,\pm1,0\big)\\
  \hline
  M^{(1)}H^{(4)}H^{(5)}M^{(8)}|\Phi_{_{\pm}}^{(\mathrm{Ei})}\rangle &
  \big(0,0,0,0,0,0,\pm1\big)\\
  \hline
\end{array}
$$
\caption{The product vectors and coordinates of vertices for
$\mathcal{W}_{_{Ei}}$.}
\renewcommand{\arraystretch}{1}
\addtolength{\arraycolsep}{3pt}
\end{table}

where
$$
\begin{array}{c}
  |\Phi_{_{\pm}}^{(\mathrm{Ei})}\rangle=|x^\pm\rangle_{_{1}}|x^+\rangle_{_{2}}...|x^+\rangle_{_{8}} \\
\end{array}
$$
Choosing any seven points among the above vertices give the
boundary half-spaces surrounding the feasible region as follows
\begin{equation}\label{ineeight}
    \begin{array}{c}
 |P_{_{1}}+(-1)^{i_{_{1}}}
 P_{_{2}}+(-1)^{i_{_{2}}}P_{_{3}}+(-1)^{i_{_{3}}}P_{_{4}}+(-1)^{i_{_{4}}}P_{_{5}}
 +(-1)^{i_{_{5}}}P_{_{1,2,3}}+(-1)^{i_{_{6}}}P_{_{1,2,4}}|\leq1 \\
 \quad,\quad\forall\;(i_{1},i_{2},...,i_{6})\in\{0,1\}^{6} \\
\end{array}
\end{equation}
We prove only the following inequality since the proof of the
other inequalities of (\ref{ineeight}) is similar to this one.
$$
 |P_{_{1}}+P_{_{2}}+P_{_{3}}+P_{_{4}}+P_{_{5}}+P_{_{1,2,3}}+P_{_{1,2,4}}|=
$$
$$
 |\alpha_{_{1}}\alpha_{_{2}}...\alpha_{_{8}}+\gamma_{_{1}}\gamma_{_{2}}...\gamma_{_{8}}+
 \gamma_{_{1}}\alpha_{_{2}}\beta{_{3}}\gamma_{_{5}}\gamma_{_{6}}\beta{_{7}}+\gamma_{_{2}}
 \gamma_{_{3}}\alpha_{_{4}}\alpha_{_{5}}\beta{_{6}}\beta{_{7}}+\alpha_{_{1}}\alpha_{_{2}}
\gamma_{_{4}}\beta{_{5}}\beta{_{6}}\gamma_{_{7}}+\alpha_{_{1}}\gamma_{_{2}}\beta{_{4}}\alpha_{_{5}}
\alpha_{_{6}}\beta{_{8}}+\beta{_{1}}\alpha_{_{2}}\alpha_{_{3}}\gamma_{_{4}}\gamma_{_{5}}\beta{_{8}}|
 \leq
$$
$$
|\alpha_{_{2}}||\alpha_{_{1}}\alpha_{_{3}}...\alpha_{_{8}}+\gamma_{_{1}}\beta{_{3}}
\gamma_{_{5}}\gamma_{_{6}}\beta{_{7}}+\alpha_{_{1}}\gamma_{_{4}}\beta{_{5}}\beta{_{6}}
\gamma_{_{7}}+\beta{_{1}}\alpha_{_{3}}\gamma_{_{4}}\gamma_{_{5}}\beta{_{8}}|+
|\gamma_{_{2}}||\gamma_{_{1}}\gamma_{_{3}}...\gamma_{_{8}}+\gamma_{_{3}}\alpha_{_{4}}
\alpha_{_{5}}\beta{_{6}}\beta{_{7}}+\alpha_{_{1}}\beta{_{4}}\alpha_{_{5}}\alpha_{_{6}}\beta{_{8}}|\leq
$$
$$
|\alpha_{_{2}}|\big(|\alpha_{_{1}}|(\alpha_{_{4}}^2+\gamma_{_{4}}^2)^{\frac{1}{2}}(\alpha_{_{3}}^2
\alpha_{_{5}}^2...\alpha_{_{8}}^2+\beta{_{5}}^2\beta{_{6}}^2\gamma_{_{7}}^2)^{\frac{1}{2}}+|\gamma_{_{1}}\beta{_{3}}
\gamma_{_{5}}\gamma_{_{6}}\beta{_{7}}|+|\beta{_{1}}\alpha_{_{3}}\gamma_{_{4}}\gamma_{_{5}}\beta{_{8}}|\big)+
$$
$$
|\gamma_{_{2}}|\big(\alpha_{_{4}}^2+\beta{_{4}}^2+\gamma_{_{4}}^2\big)^{\frac{1}{2}}
\big(\gamma_{_{1}}^2\gamma_{_{3}}^2\gamma_{_{5}}^2...\gamma_{_{8}}^2+\gamma_{_{3}}^2
\alpha_{_{5}}^2\beta{_{6}}^2\beta{_{7}}^2+\alpha_{_{1}}^2\alpha_{_{5}}^2\alpha_{_{6}}^2\beta{_{8}}^2\big)
^{\frac{1}{2}}\leq
$$
$$
|\alpha_{_{2}}|\big(\alpha_{_{1}}^2+\beta{_{1}}^2+\gamma_{_{1}}^2\big)^{\frac{1}{2}}\big(\alpha_{_{3}}^2
\alpha_{_{5}}^2...\alpha_{_{8}}^2+\beta{_{5}}^2\beta{_{6}}^2\gamma_{_{7}}^2+\beta{_{3}}^2
\gamma_{_{5}}^2\gamma_{_{6}}^2\beta{_{7}}^2+\alpha_{_{3}}^2\gamma_{_{4}}^2\gamma_{_{5}}^2\beta{_{8}}^2\big)
^{\frac{1}{2}}+
$$
$$
|\gamma_{_{2}}|\big(\gamma_{_{1}}^2\gamma_{_{3}}^2\gamma_{_{5}}^2...\gamma_{_{8}}^2+\gamma_{_{3}}^2
\alpha_{_{5}}^2\beta{_{6}}^2\beta{_{7}}^2+\alpha_{_{1}}^2\alpha_{_{5}}^2\alpha_{_{6}}^2\beta{_{8}}^2\big)
^{\frac{1}{2}}\leq
\big(\alpha_{_{2}}^2+\gamma_{_{2}}^2\big)^{\frac{1}{2}}\ \times
$$
$$
\big(\alpha_{_{3}}^2\alpha_{_{5}}^2...\alpha_{_{8}}^2+\beta{_{5}}^2\beta{_{6}}^2\gamma_{_{7}}^2+\beta{_{3}}^2
\gamma_{_{5}}^2\gamma_{_{6}}^2\beta{_{7}}^2+\alpha_{_{3}}^2\gamma_{_{4}}^2\gamma_{_{5}}^2\beta{_{8}}^2+
\gamma_{_{1}}^2\gamma_{_{3}}^2\gamma_{_{5}}^2...\gamma_{_{8}}^2+\gamma_{_{3}}^2
\alpha_{_{5}}^2\beta{_{6}}^2\beta{_{7}}^2+\alpha_{_{1}}^2\alpha_{_{5}}^2\alpha_{_{6}}^2\beta{_{8}}^2\big)
^{\frac{1}{2}}\leq
$$
$$
\big[\alpha_{_{5}}^2\big(\alpha_{_{6}}^2(\alpha_{_{3}}^2\alpha_{_{7}}^2\alpha_{_{8}}^2+
\alpha_{_{1}}^2\beta{_{8}}^2)+\gamma_{_{3}}^2\beta{_{6}}^2\beta{_{7}}^2\big)+
\beta{_{5}}^2\beta{_{6}}^2\gamma_{_{7}}^2+\gamma_{_{5}}^2\big(\beta{_{3}}^2\gamma_{_{6}}^2\beta{_{7}}^2+
\alpha_{_{3}}^2\gamma_{_{4}}^2\beta{_{8}}^2+
\gamma_{_{1}}^2\gamma_{_{3}}^2\gamma_{_{6}}^2\gamma_{_{7}}^2\gamma_{_{8}}^2\big)\big]^{\frac{1}{2}}\leq
$$
$$
\big(\alpha_{_{5}}^2+\beta{_{5}}^2+\gamma_{_{5}}^2\big)^{\frac{1}{2}}\leq1
$$
where, we have used the Cauchy-Schwartz inequality and the
equations (\ref{exval}) and (\ref{exval1}) repeatedly.
\par
Inequalities obtained from putting the vertex points are
$$
\begin{array}{c}
 a_{_{0}} \geq |a_{_{i}}|\quad\quad i=1,...,5 \quad,\quad a_{_{0}} \geq |a_{_{1,2,3}}|
 \quad,\quad a_{_{0}} \geq |a_{_{1,2,4}}|  \\
\end{array}
$$
{\bf Nine-qubit SEWs}\\
 Consider the following Hermitian operator
$$
\mathcal{W}_{_{\mathrm{Ni}}}=a_{_{0}}I_{_{2^9}}+\sum_{i=1}^{8}
a_{_{i}}S_{_{i}}^{(\mathrm{Ni})}+
a_{_{1,3}}S_{_{1}}^{(\mathrm{Ni})}S_{_{3}}^{(\mathrm{Ni})}
$$
Eigenvalues of $\mathcal{W}_{_{Ni}}$ are
$$
a_{_{0}}+ \sum_{j=1}^{8}(-1)^{i_{j}} a_{_{j}}+(-1)^{i_{1}+i_{3}}
a_{_{1,3}}\quad\quad
\forall\;(i_{1},i_{2},...,i_{8})\in\{0,1\}^{8}
$$
The vertex points of feasible region are listed in table 13

\begin{table}[h]
\renewcommand{\arraystretch}{1.2}
\addtolength{\arraycolsep}{0pt}
$$
\begin{array}{|c|c|}\hline
 \mathrm{Product\ state} & (P_{_{1}},P_{_{2}},P_{_{3}},P_{_{4}}, P_{_{5}},P_{_{6}}, P_{_{7}},P_{_{8}},P_{_{1,3}})\\
  \hline
  \Lambda_{_{i_{1},i_{2}}}^{(\mathrm{Ni})}|\Phi^{(\mathrm{Ni})}\rangle & \big((-1)^{i_{1}},(-1)^{i_{2}},
  0,0,0,0,0,0,0\big) \\
   \hline
  {\Lambda}_{_{i_{1},i_{2},i_{3}}}^{(\mathrm{Ni})}|\Phi^{(\mathrm{Ni})}\rangle & \big((-1)^{i_{1}},0,
  0,0,0,0,(-1)^{i_{2}},(-1)^{i_{3}},0\big)\\
  \hline
  {\Lambda'}_{_{i_{1},i_{2},i_{3}}}^{(\mathrm{Ni})}|\Phi^{(\mathrm{Ni})}\rangle & \big(0,(-1)^{i_{1}},(-1)^{i_{2}}
  ,(-1)^{i_{3}},0,0,0,0,0\big)\\
  \hline
  {\Lambda''}_{_{i_{1},i_{2},i_{3}}}^{(\mathrm{Ni})}|\Phi^{(\mathrm{Ni})}\rangle &
  \big(0,0,0,0,0,0,(-1)^{i_{1}},(-1)^{i_{2}},(-1)^{i_{3}}\big)\\
  \hline
 {\Lambda}_{_{i_{1},i_{2},i_{3},i_{4},i_{5},i_{6}}}^{(\mathrm{Ni})}|\Phi^{(\mathrm{Ni})}\rangle
&
  \big(0,0,(-1)^{i_{1}},(-1)^{i_{2}},(-1)^{i_{3}},(-1)^{i_{4}},(-1)^{i_{5}},(-1)^{i_{6}},0\big)\\
  \hline
\end{array}
$$
\caption{The product vectors and coordinates of vertices for
$\mathcal{W}_{_{Ni}}$.}
\renewcommand{\arraystretch}{1}
\addtolength{\arraycolsep}{3pt}
\end{table}
where
$$
\begin{array}{c}
    |\Phi^{(\mathrm{Ni})}\rangle=|x^+\rangle_{_{1}}|x^+\rangle_{_{2}}...|x^+\rangle_{_{9}} \\
    \Lambda_{_{i_{1},i_{2}}}^{(\mathrm{Ni})}=\big(\sigma_{z}^{(1)}\big)^{i_{_{1}}}\big(\sigma_{z}^{(7)}\big)^{i_{_{2}}} \\
    {\Lambda}_{_{i_{1},i_{2},i_{3}}}^{(\mathrm{Ni})}=\big(\sigma_{z}^{(1)}\big)^{i_{_{1}}}\big(\sigma_{x}^{(7)}\big)^{i_{_{2}}}
    \big(\sigma_{x}^{(9)}\big)^{i_{_{3}}}H^{(7)}H^{(8)}H^{(9)}\\
    {\Lambda'}_{_{i_{1},i_{2},i_{3}}}^{(\mathrm{Ni})}=\big(\sigma_{x}^{(1)}\big)^{i_{_{1}}}\big(\sigma_{x}^{(3)}\big)^{i_{_{2}}}
    \big(\sigma_{z}^{(4)}\big)^{i_{_{3}}}H^{(1)}H^{(2)}H^{(3)} \\
    {\Lambda''}_{_{i_{1},i_{2},i_{3}}}^{(\mathrm{Ni})}=\big(\sigma_{z}^{(1)}\big)^{i_{_{1}}}\big(\sigma_{x}^{(7)}\big)^{i_{_{2}}}
    \big(\sigma_{x}^{(9)}\big)^{i_{_{3}}}(M^{(1)})^\dagger M^{(2)} \\
    {\Lambda}_{_{i_{1},i_{2},i_{3},i_{4},i_{5},i_{6}}}^{(\mathrm{Ni})}=\big(\sigma_{x}^{(1)}\big)^{i_{_{1}}}
    \big(\sigma_{x}^{(3)}\big)^{i_{_{2}}}\big(\sigma_{x}^{(4)}\big)^{i_{_{3}}}\big(\sigma_{x}^{(6)}\big)^{i_{_{4}}}
    \big(\sigma_{x}^{(7)}\big)^{i_{_{5}}}\big(\sigma_{x}^{(9)}\big)^{i_{_{6}}}\bigotimes_{j=1}^9H^{(j)} \\
 \end{array}
$$
which in all of the above operators we assume that
$(i_{1},...,i_{j})\in\{0,1\}^{j},\ \mathrm{with} \ j=2,3,6$ . By
choosing any eight points among the above vertices give the
half-spaces surrounding the feasible region as follows
$$
\begin{array}{c}
|P_{_{1}}+P_{_{i}}\pm P_{_{1,3}}|\leq 1\quad,\quad
|P_{_{1}}-P_{_{i}}\pm P_{_{1,3}}|\leq 1\quad,\quad i=3,4,5,6 \\
|P_{_{2}}\pm P_{_{j}}|\leq 1\quad,\quad j=7,8 \\
\end{array}
$$
The proof of the above inequalities are straight forward.
Inequalities obtained from putting the vertex points are
$$
\begin{array}{c}
 a_{_{0}} \geq |a_{_{1}}|+|a_{_{2}}|\quad,\quad a_{_{0}} \geq |a_{_{1}}|+|a_{_{7}}|+|a_{_{8}}| \\
 a_{_{0}} \geq |a_{_{2}}|+|a_{_{3}}|+|a_{_{4}}|\quad,\quad a_{_{0}}\geq \sum_{j=3}^{8}
|a_{_{j}}|\quad,\quad a_{_{0}} \geq |a_{_{7}}|+|a_{_{8}}|+|a_{_{1,3}}| \\
\end{array}
$$

\newpage
{\bf Figure Captions}

 {\bf Figure-1:} 8-simplex displaying the
feasible region of the two-qubit GHZ SEW.

{\bf Figure-2:} Graphs corresponding to different graph states
where the first two ones are graph states and the others are graph
codes. (a) The star graph describing a GHZ state. (b) The linear
graph describing a cluster state.  The graph codes for (c)
five-qubit , (d) seven-qubit, (e) eight-qubit and (f) nine-qubit
stabilizer groups.

\end{document}